\documentclass[twocolumn, times]{aastex62}

\accepted{\today}
\submitjournal{AJ}

\shorttitle{Jupiter's Mesoscale Waves at 5 $\mu$m}
\shortauthors{Fletcher et al.}
\begin{document}

%\begin{frontmatter}



\title{Jupiter's Mesoscale Waves Observed at 5 $\mu$m by Ground-Based Observations and Juno JIRAM}


\correspondingauthor{Leigh Fletcher}
\email{leigh.fletcher@le.ac.uk}

\author{Leigh N. Fletcher}
\affil{Department of Physics and Astronomy, University of Leicester, University Road, Leicester, LE1 7RH, UK.}

\author{H. Melin}
\affil{Department of Physics and Astronomy, University of Leicester, University Road, Leicester, LE1 7RH, UK.}

\author{A. Adriani}
\affil{INAF-Istituto di Astrofisica e Planetologia Spaziali, Roma, Italy.}

\author{A.A. Simon}
\affiliation{NASA Goddard Space Flight Center Solar System Exploration Division (690) Greenbelt, MD 20771, USA}

\author{A. Sanchez-Lavega}
\affiliation{Departamento de Física Aplicada I, Escuela de Ingeniería de Bilbao, UPV/EHU, Plaza Ingeniero Torres Quevedo, 1, 48013 Bilbao, Spain.}

\author{P.T. Donnelly}
\affil{Department of Physics and Astronomy, University of Leicester, University Road, Leicester, LE1 7RH, UK.}

\author{A. Antu\~{n}ano}
\affil{Department of Physics and Astronomy, University of Leicester, University Road, Leicester, LE1 7RH, UK.}

\author{G.S. Orton}
\affiliation{Jet Propulsion Laboratory, California Institute of Technology, 4800 Oak Grove Drive, Pasadena, CA 91109, USA}

\author{R. Hueso}
\affiliation{Departamento de Física Aplicada I, Escuela de Ingeniería de Bilbao, UPV/EHU, Plaza Ingeniero Torres Quevedo, 1, 48013 Bilbao, Spain.}

\author{E. Kraaikamp}
\affiliation{Jourdanstraat 121/8, 1060, Sint-Gillis, Belgium.}

\author{M.H. Wong}
\affiliation{University of California at Berkeley, Astronomy Department Berkeley, CA 947200-3411, USA.}
\author{M. Barnett}
\affiliation{University of California at Berkeley, Astronomy Department Berkeley, CA 947200-3411, USA.}

\author{M.L. Moriconi}
\affil{CNR-Istituto di Scienze dell Atmosfera e del Clima, Bologna e Roma, Italy.}
\author{F. Altieri}
\affil{INAF-Istituto di Astrofisica e Planetologia Spaziali, Roma, Italy.}

\author{G. Sindoni}
\affil{INAF-Istituto di Astrofisica e Planetologia Spaziali, Roma, Italy.}

\begin{abstract}

We characterise the origin and evolution of a mesoscale wave pattern in Jupiter's North Equatorial Belt (NEB), detected for the first time at 5 $\mu$m using a 2016-17 campaign of `lucky imaging' from the VISIR instrument on the Very Large Telescope and the NIRI instrument on the Gemini observatory, coupled with M-band imaging from Juno's JIRAM instrument during the first seven Juno orbits.  The wave is compact, with a $1.1-1.4^\circ$ longitude wavelength (wavelength 1,300-1,600 km, wavenumber 260-330) that is stable over time, with wave crests aligned largely north-south between $14$ and $17^\circ$N (planetographic).  The waves were initially identified in small ($10^\circ$ longitude) packets immediately west of cyclones in the NEB at $16^\circ$N, but extended to span wider longitude ranges over time.  The waves exhibit a 7-10 K brightness temperature amplitude on top of a $\sim210$-K background at 5 $\mu$m.  The thermal structure of the NEB allows for both inertio-gravity waves and gravity waves.  Despite detection at 5 $\mu$m, this does not necessarily imply a deep location for the waves, and an upper tropospheric aerosol layer near 400-800 mbar could feature a gravity wave pattern modulating the visible-light reflectivity and attenuating the 5-$\mu$m radiance originating from deeper levels.  Strong rifting activity appears to obliterate the pattern, which can change on timescales of weeks.  The NEB underwent a new expansion and contraction episode in 2016-17 with associated cyclone-anticyclone formation, which could explain why the mesoscale wave pattern was more vivid in 2017 than ever before.

\end{abstract}


%\linenumbers

%%%%%%%%%%%%%%%%%%%%%%%%%%%%%%%%%%%%%%%%%%%%%%
%%%%%%%%%%%%%%%%%%%%%%%%%%%%%%%%%%%%%%%%%%%%%%
%%%%%%%%%%%%%%%%%%%%%%%%%%%%%%%%%%%%%%%%%%%%%%
\section{Introduction}
\label{intro}

Jupiter's 5-$\mu$m window is a unique region of the infrared spectrum where a dearth of gaseous opacity permits deep observations of Jupiter's cloud-forming region, below the stably-stratified upper troposphere \citep[e.g.,][]{77terrile}.  Although reflected sunlight from upper tropospheric clouds and hazes contributes in Jupiter's zones, the dominant contribution in the belts is from thermal emission in the 4-8 bar range, with clouds appearing in silhouette against the bright background of the warm troposphere.  The window is bounded at long wavelengths by the $2\nu_2$ and $\nu_4$ absorption bands of NH$_3$, and at short wavelengths by phosphine absorption in the broad $\nu_2+\nu_4$ band between 4.69-4.78 $\mu$m.  A host of additional disequilibrium species (AsH$_3$, GeH$_4$, CO, etc.) contribute to the spectrum, but these are all modulated by opacity variations in aerosol layers - both the upper tropospheric clouds related to NH$_3$ ice formation, and the mid-tropospheric clouds associated with the formation of NH$_4$SH \citep{15giles,17giles}.  H$_2$O ice may also contribute opacity in certain regions \citep{15bjoraker}.  Combined, this makes the 5-$\mu$m window useful as a diagnostic of the dynamics and circulation of the cloud-forming region.  

Numerous studies have exploited this window in the past decade, ranging from space-based observations at low-spectral resolution by Cassini VIMS \citep{15giles} and Juno JIRAM \citep{17grassi}; to ground-based spectra at high spectral resolution by the VLT \citep{15giles, 17giles, 17giles_nh3}, Gemini \citep{15bjoraker} and the IRTF \citep{16fletcher_texes}; and narrow-band photometric imaging from the IRTF, Gemini, and Keck.  Thermal imaging has been used to explore the dynamics of the Great Red Spot \citep{10fletcher_grs}; the bright peripheral rings around other vortices \citep{10depater_jup}; and the changes in aerosol opacity associated with Jupiter's belt/zone fades and revivals \citep{11fletcher_fade, 17fletcher_seb}.  Observations at 5 $\mu$m were therefore deemed to be an essential component of the Earth-based supporting campaign for NASA's Juno mission \citep{17bolton}, both preceding its arrival in July 2016 and during each of its 53.5-day polar orbits. In this work, we report on the use of a new capability of ESO's Very Large Telescope's VISIR instrument \citep{04lagage}, allowing it to perform `lucky imaging' of Jupiter in burst mode to `freeze' the seeing and provide diffraction-limited performance at 5 $\mu$m.  This removes the need for adaptive optics using the Galilean satellites for wavefront sensing \citep{10depater_jup}, therefore increasing the flexibility of observations so that they could be designed to coincide with each of Juno's perijoves.  These are supplemented by Juno JIRAM M-band images with partial spatial coverage in 2016-17, and by an additional campaign of M-band lucky imaging with the NIRI instrument on Gemini North.

The VLT images provide near full-disc coverage of Jupiter at spatial resolutions of 0.15", equating to horizontal resolutions of $\sim470$ km at opposition ($0.39^\circ$ longitude at the equator) and $\sim570$ km ($0.46^\circ$ longitude) at quadrature. At this resolution, it is possible to resolve structures less than a degree of longitude in width, providing 5-$\mu$m access to length scales typical of the Hubble Space Telescope \citep{18simon} and the best amateur imagers.  In particular, we reach the $\sim1400$-km length scale of a wave pattern in Jupiter's North Equatorial Belt (NEB) near $16^\circ$N (planetographic latitude) that was previously reported by Voyager 2 \citep{79smith} and later rediscovered in Hubble imaging in January 2015 \citep{15simon}.  \citet{15simon} suggested that these `mesoscale' waves were formed via a baroclinic instability mechanism potentially associated with cyclogenesis, but the rarity of their presence proved a challenge to more detailed analysis.  The VLT 5-$\mu$m imaging coincided with their dramatic reappearance in 2016-17, at a time when the NEB was undergoing an expansion and contraction episode as part of its 4-5 year cycle \citep{17fletcher_neb}.  In two companion articles, \citet{18simon} report the properties of the NEB wave as observed in the visible range by Hubble and amateur ground-based imaging, and \citet{18adriani} reports on Juno spectroscopy of the wave pattern.  Section \ref{data} describes the lucky-imaging process used to acquire the VLT data and reveals that this wave is detectable at 5 $\mu$m; Section \ref{res} describes the chronology of the mesoscale wave pattern and its association with NEB activity; and Section \ref{discuss} explores the implications of the 5-$\mu$m detection and the different potential wave mechanisms responsible for this pattern. 

\section{Data Acquisition}
\label{data}

\subsection{VISIR Observations}

The  VLT Imager and Spectrometer for the mid-infrared \citep[VISIR,][]{04lagage} returned to the 8.2-m UT3/Melipal telescope in 2015, after a 3-year refurbishment.  The new $1024\times1024$ Raytheon Aquarius IBC detector offers a pixel size of 0.0453"/pixel over a $38\times38$" field of view, smaller than the disc of Jupiter at opposition, but sufficient to oversample the 0.15" spatial resolution of the 5-$\mu$m imaging.  Following interactions with Hans-Ulrich Kaufl in March 2015, an M-band filter was located and inserted into VISIR.  Given the high sky background, the individual integration times are usually a few milliseconds, and the upgraded VISIR now offers a new burst mode, whereby all individual detector exposures are recorded, rather than just retaining the averages per nodding cycle.  We were awarded Science Verification observations in Period 96 (February 2016), immediately prior to Juno's arrival at Jupiter, to test this new M-band `lucky-imaging' capability. Once validated, this became a part of our regular observing sequence in Period 98 (December 2016 onwards).  We targeted all opportunities when Jupiter was available for more than one hour within a week of Juno's perijove encounters.  Unfortunately, weather constraints at Paranal and scheduling competition for service-mode observations restricted the dataset to six distinct epochs, as shown in Table \ref{tab:data}.  The epochs were primarily clustered during the 2016-17 apparition (centred on Jupiter's opposition on April 7th 2017).  Regular observations at 7-20 $\mu$m (i.e., without burst mode) demonstrated the changes associated with the NEB expansion during this period \citep{17fletcher_neb}.

%\begin{figure*}
%\begin{centering}
%\includegraphics[angle=0,scale=0.8]{.pdf}
%\caption{}
%\label{}
%\end{centering}
%\end{figure*}

\begin{table*}
\caption{Observations at 5 $\mu$m presented in this work.  The System III Longitudes span from east to west.}
\centering
\begin{tabular}{c c c c c}
\hline
Date & Observations & Number of Obs. & ID & Comments \\
\hline
2016-02-26 & 05:48-06:20 & 4 & 60.A-9620 & Science Verification, no waves $30-160^\circ$W \\
2016-08-27 & - & - & JIRAM PJ1 & Waves $36-52^\circ$W \\
2016-12-18 & 08:21-08:25 & 1 & 098.C-0681& Waves $80-105^\circ$W\\
2017-01-11 & 08:29-08:33 & 1 & 098.C-0681& Waves $95-117^\circ$W \\
2017-02-02 & - & - & JIRAM PJ4 & Waves $68-82^\circ$W \\
2017-02-05 & 14:13--15:48 & 62 & GN-2017A-Q-60 & Waves $44-70^{\circ}$W + $165-193^{\circ}$W\\
2017-02-06 & 05:45-05:49 & 1 & 098.C-0681 & No waves $230-360^\circ$W\\ 
2017-03-16 & 05:41-05:46 & 1 & 098.C-0681& No waves $200-320^\circ$W\\
2017-03-27 & - & - & JIRAM PJ5 & Waves $40-85^\circ$W \\
2017-05-19 & - & - & JIRAM PJ6 & No waves $205-260^\circ$W \\
2017-07-11 & - & - & JIRAM PJ7 & Waves $280-30^\circ$W \\
2017-07-20 & 01:04-01:17 & 1 & 099.C-0612 & No waves $280-50^\circ$W (poor seeing)\\
%2018-02-20 & 08:45-08:52 & 1 & 0100.C-0186 & Juno PJ11; GRS on dusk limb.\\
\hline
\label{tab:data}
\end{tabular}
\end{table*}

A single observing block has $N_{nod}$ nodding cycles (usually four or five), each producing two files:  the first performed chopping on Jupiter itself, the second performed chopping at a position 25 arcseconds away.  The observations were designed such that the direction of the chop was determined by the particular hemisphere that we were focussing on:  chopping the telescope north if we desired unobstructed views of the northern hemisphere; chopping the telescope south if we wanted unobstructed views of the southern hemisphere.  The total integration time, divided by the number of nodding cycles and the 3-4 Hz chopping frequency determined the number of chopping cycles within each individual file.  The individual integration time of each frame was 11.4-20.8 milliseconds, depending on the chopping frequency used.  We took the sum of the chop-nod position which encompassed only blank sky, and subtracted this from every frame of the chop-nod position that targeted Jupiter alone (effectively discarding 50\% of the data).  This produced hundreds of individual frames for each nodding cycle.  These were saved as both MP4 files for quick inspection and uncompressed TIFF files for further processing.  The movies show Jupiter moving around on the detector and coming in and out of focus as the seeing varies over millisecond timescales.

The TIFF files were imported into AutoStakkert\footnote{autostakkert.com}, a software tool developed by E. Kraaikamp to identify and stack those frames with the best image quality \citep{16kraaikamp}.  The software centres and aligns the individual frames, using a `surface' mode to track individual bright features within an alignment box on Jupiter throughout the sequence.  The frame quality is estimated via measurements of local gradients from frame to frame, and used to rank the frames from best to worst.  One large alignment point, encompassing the NEB and SEB of Jupiter, was then used as the anchor to stack the top 2, 5, 10 and 20\% of frames.  As more frames sometimes worsened the quality of the stack, we made a qualitative selection of one stack for continued analysis.  The limb of the planet was fitted to assign latitudes, longitudes and emission angles to each pixel, which were then projected onto a $0.25^\circ$-resolution cylindrical map.  No attempt was made to radiometrically calibrate the data.  
\subsection{Gemini Observations}
The VISIR data were supplemented by a single mosaic 5-$\mu$m image from the Near InfraRed Imager and Spectrometer \citep[NIRI,][]{03hodapp} on Gemini-North, using a similar lucky imaging process without adaptive optics.  NIRI's $1024\times1024$ pixel InSb array and 0.0218"/pixel scale limits the field of view to $22\times22$", which was mosaicked across the disc to generate a map on February 5th 2017 (Table \ref{tab:data}).  Details of the wider Gemini/NIRI programme will be described in a forthcoming paper.

\subsection{Juno JIRAM Observations}

The ground-based observations are compared to partial maps of the NEB acquired by the JIRAM instrument on the Juno spacecraft.  The JIRAM imager and spectrometer \citep{17adriani} is mounted on the Juno spacecraft and has been operating since August 2016.  Juno approaches the planet every 53 days with different attitudes according to primary science objectives \citep{17bolton}.  In the first part of the mission, most of the flybys were favourable to JIRAM and the planet could be observed during the approach with good views of the northern hemisphere. The NEB was observed with variable coverage during the orbit 1, 4, 5, 6 and 7 (during orbit 2 and 3 the instrument was not operating).  The JIRAM imager has a channel working around 5 $\mu$m wavelength. Single images have a size of $432\times128$ pixels. Pixels are square-shaped with an angular resolution of $240\times240$ $\mu$m, which defines the spatial resolution at the cloud level as a function of the distance of the spacecraft from the planet. The images reported hereafter present an average spatial resolution of about 250 km at the cloud level, twice as good as the VLT resolution reported above. When the spacecraft's spinning plane intersects the planet, JIRAM can make scans from south to north, changing pointing approximately every 30 s (spacecraft spinning period). Maps of a limited range of latitudes can be built by mosaicking images from different observing sequences, although some artefacts occur because the atmosphere has evolved during the interval between sequences. Dates of the JIRAM data acquisitions are reported in Table \ref{tab:data}.

\section{Results}
\label{res}
%\subsection{Jupiter Activity in 2016-17}

\begin{figure*}
\begin{centering}
\includegraphics[angle=0,width=1.0\textwidth]{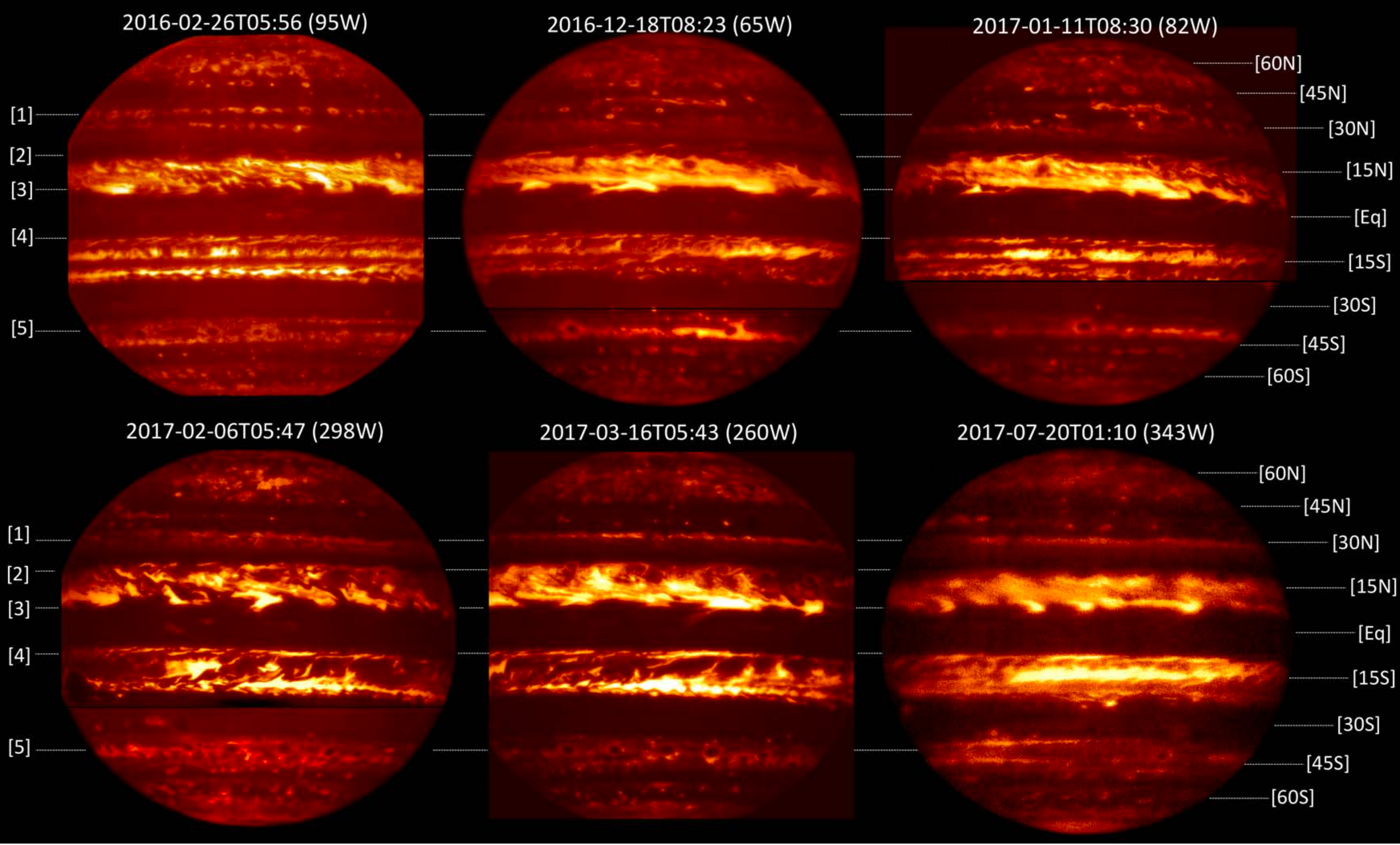}
\caption{Six examples of VISIR Burst-Mode imaging at 5 $\mu$m.  The December 2016, January 2017 and February 2017 images were assembled from two sets of images, one targeting the north, one targeting the south.  Horizontal numbered lines indicate regions discussed in the main text and appendix, and a latitude scale is included on the right-hand side.  The central meridian longitude is indicated in each frame.   }
\label{globes}
\end{centering}
\end{figure*}

\begin{figure*}
\begin{centering}
\includegraphics[angle=0,width=1.0\textwidth]{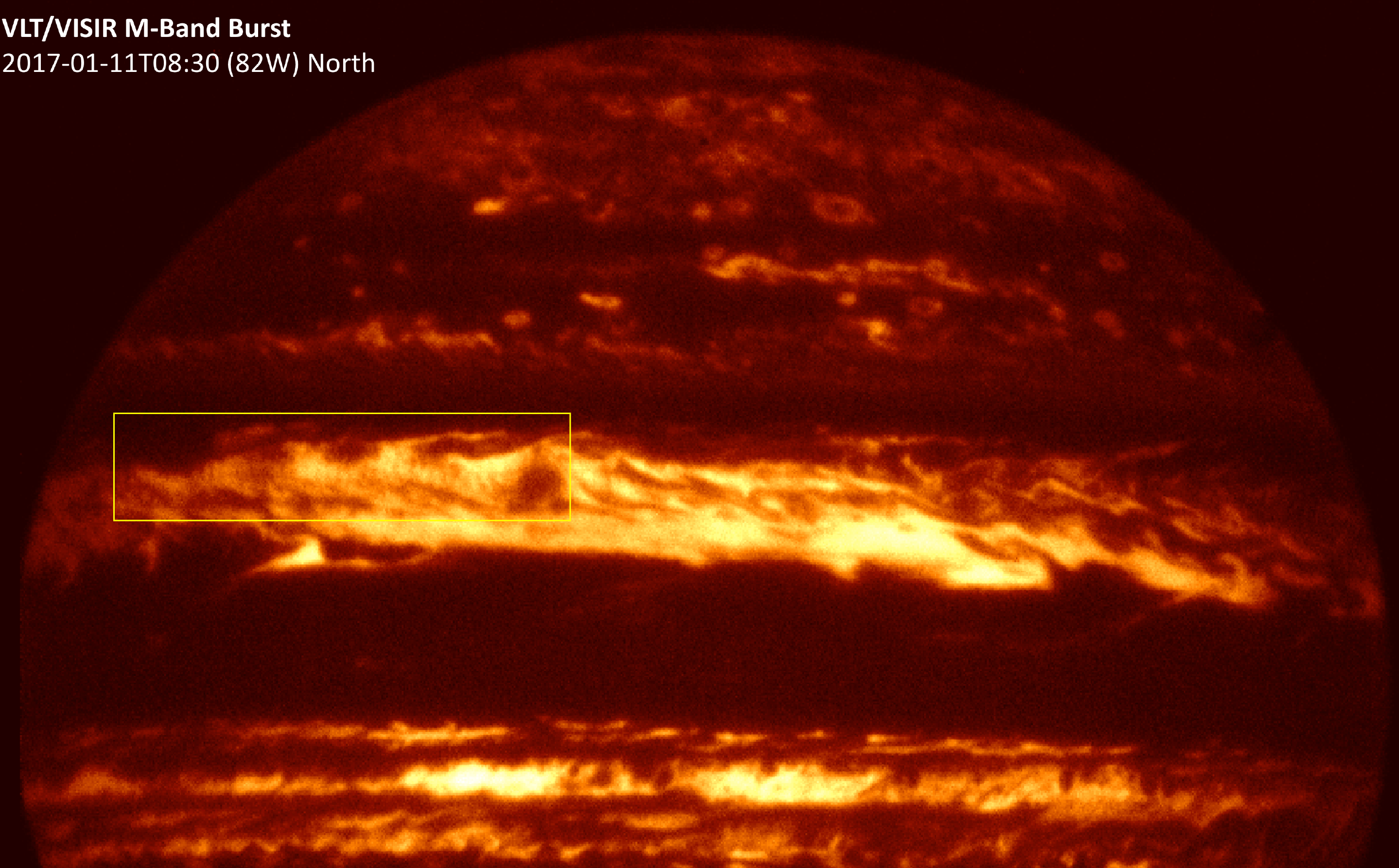}
\caption{Close-up view of the VLT 5-$\mu$m observations in January 2017, highlighting the mesoscale waves to the west of the dark cyclone B1.  }
\label{jan_globe}
\end{centering}
\end{figure*}

Fig. \ref{globes} presents the full-disc images of Jupiter acquired using VISIR's burst mode between 2016-17, providing global context for the NEB activity.  Fig. \ref{jan_globe} provides a larger-scale example of a northern hemisphere image in January 2017, with the NEB wave pattern highlighted.  This was the first evidence that the NEB wave of \citet{15simon} was visible at 5 $\mu$m, and supported the view that it could be associated with cyclogenesis.   

%The northern and southern hemispheres are mapped separately in Figs. \ref{cmaps_north}-\ref{cmaps_south}, highlighting some of the variability observed in each hemisphere.  This provides context for the chronology of the NEB wave activity described in Section \ref{nebwaves}.  

\subsection{Status of the North Equatorial Belt (NEB) in 2016-17}

\begin{figure*}
\begin{centering}
\includegraphics[angle=0,width=1.0\textwidth]{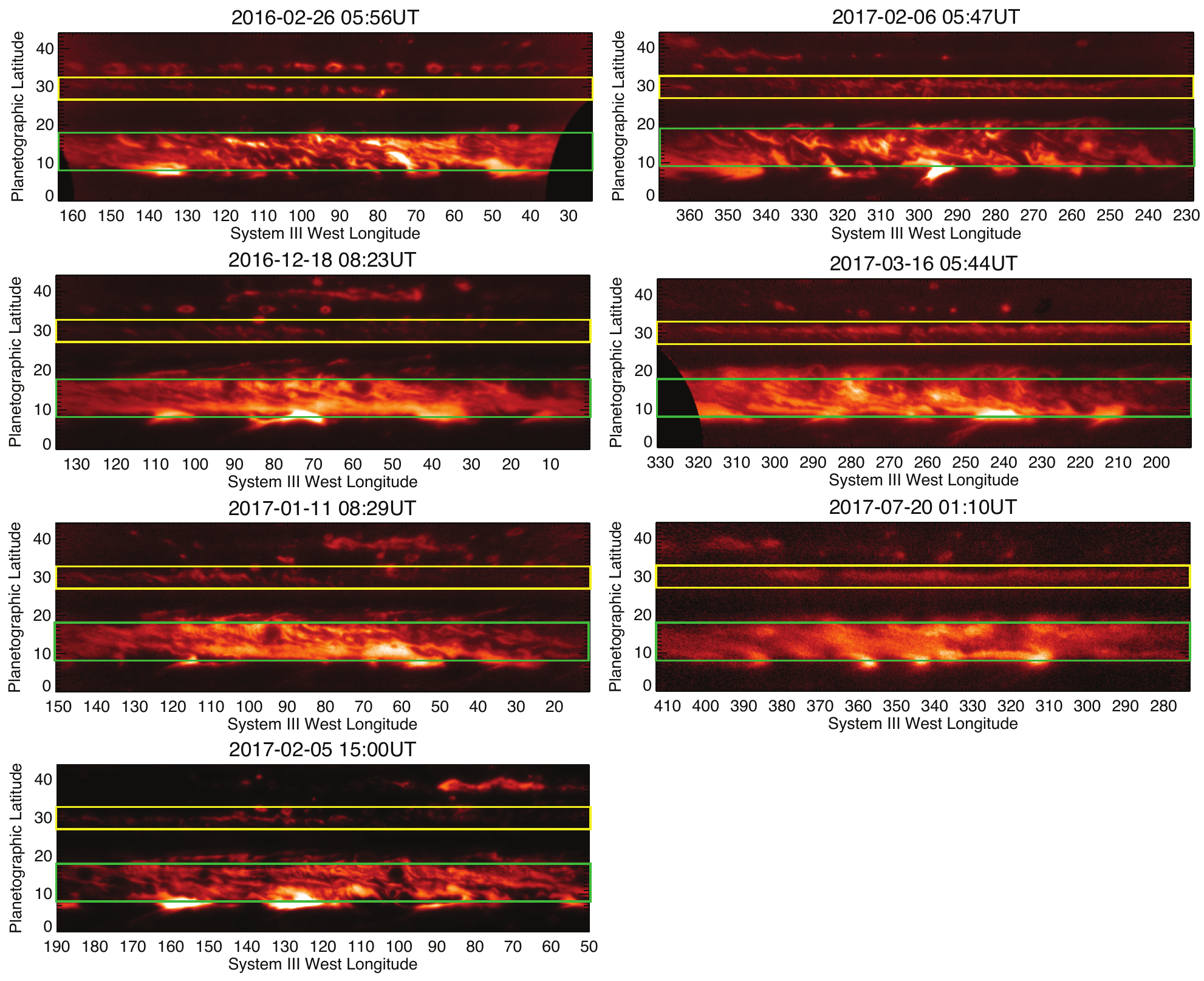}
\caption{Maps of 5-$\mu$m emission in the northern hemisphere, highlighting the North Temperate Belt (NTB) region in yellow and the North Equatorial Belt (NEB) region in green.  }
\label{cmaps_north}
\end{centering}
\end{figure*}

Fig. \ref{cmaps_north} shows the evolution of the NEB during a new phase of NEB expansion in 2017 (all latitudes are planetographic).  During these expansion events, the dark colouration of the NEB is seen to extend from $17^\circ$N to $20^\circ$N, encroaching on the typically-white North Tropical Zone (NTrZ) - region [2] of Fig. \ref{globes}.   This dark colouration coincides with warmer temperatures and bright 5-$\mu$m emission, implying the removal of the white NTrZ aerosols during an expansion.  The first phase of the expansion in 2015-16 was documented by \citet{17fletcher_neb}, but had stalled and regressed by the time of Juno's arrival in July 2016, and never spanned all longitudes.  At its peak, the expanded region spanned $\sim145^\circ$ of longitude west of a large anticyclone (White Oval Z, $19^\circ$N, $283^\circ$W in February 2016).  The expanded region reached an unusual cyclone-anticyclone pair, which was observed during our first VLT lucky-imaging campaign in February 2016 (Fig. \ref{feb2016}), before Juno's arrival.  The anticlockwise motion of the cyclone can be observed near $16^\circ$N, $55^\circ$W, embedded within the NEB.  This cyclone will be labelled B1 for the purposes of this study.  The anticlockwise motion of the anticyclone can seen at $19^\circ$N, $43^\circ$W, nominally within the NTrZ.  The expanded NEB had regressed at all longitudes by the time of Juno's arrival, and the NEB had its normal width in August 2016 (Fig. \ref{jiram}a) and December 2016 (Fig. \ref{cmaps_north}).

Maps in February and March 2017 (Fig. \ref{cmaps_north}) show a second phase of cloud-clearing within the NTrZ, where the 5-$\mu$m emission is visible at latitudes poleward of the NEBn jet at $17^\circ$N (planetographic).  This time, the expansion spread over all longitudes.  At the end of an expansion episode, the undulations of the northern edge of the NEB transform into a chain of cyclones near $16^\circ$N (sometimes known as barges) and anticyclonic white ovals (AWOs) near $19^\circ$N \citep{17fletcher_neb, 17rogers}.  This pattern of cyclone-anticyclone pairs can be seen in March 2017 (see Section \ref{nebwaves}).  In addition, the southern edge of the NEB (region [3] in Fig. \ref{globes}) exhibits the familiar chain of 5-$\mu$m bright hotspots that move rapidly eastward with the prograde NEBs jet at $7^\circ$N.  The influence of the NEB expansion and the presence of the NEB cyclones will be described in Section \ref{nebwaves}.

\subsection{Chronology of NEB Waves}
\label{nebwaves}

\begin{figure*}
\begin{centering}
\includegraphics[angle=0,width=0.9\textwidth]{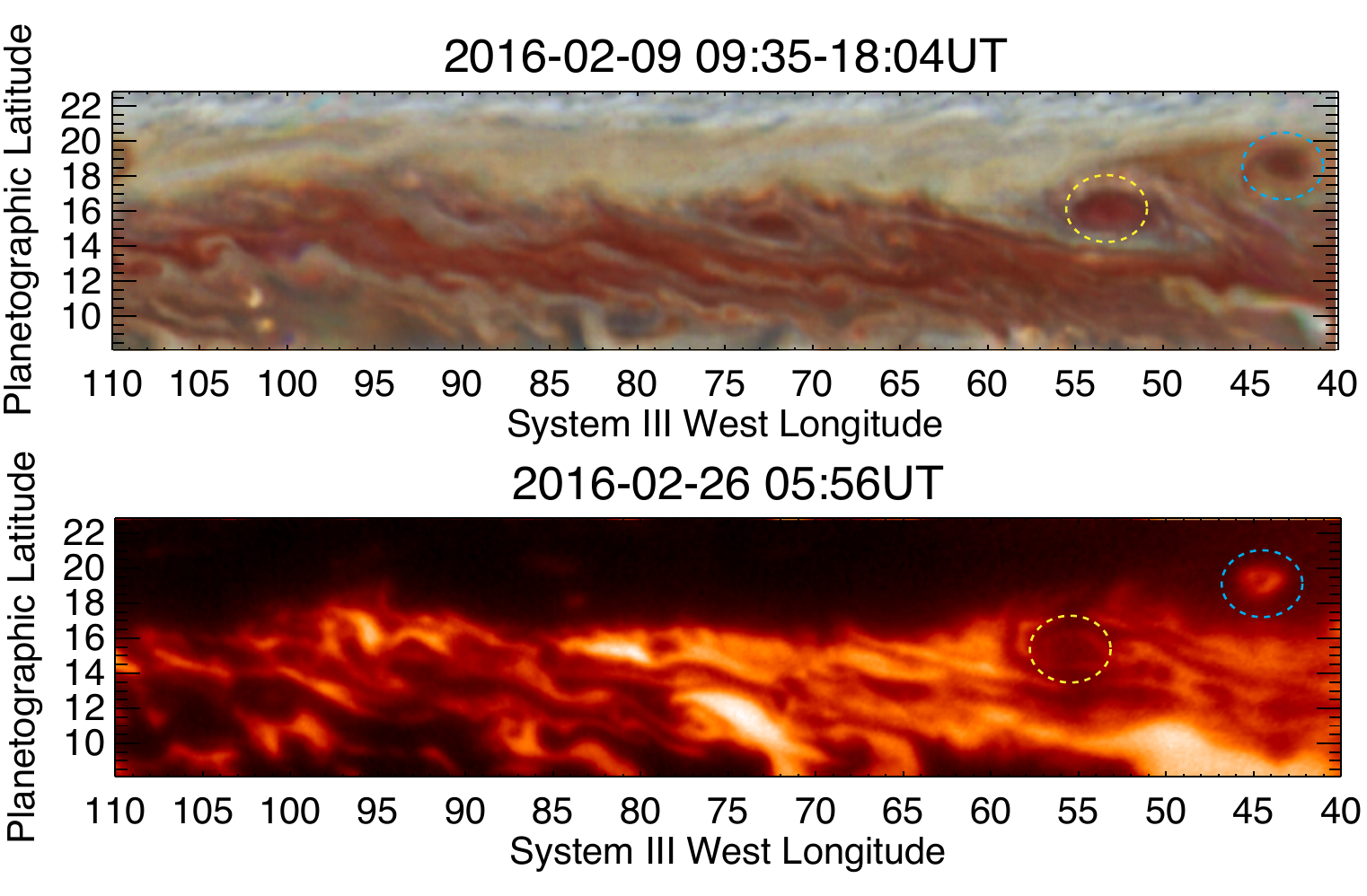}
\caption{February 26th, 2016 observation of the NEB in a non-expanded domain \citep{17fletcher_neb} compared to Hubble imaging from the OPAL programme on February 9th 2016 (17 days earlier).  Cyclones equatorward the NEBn jet are indicated by yellow ovals, anticyclones poleward of the NEBn jet by blue ovals.  Both HST and VLT images have been sharpened by means of a high-pass filter.  No waves are detectable in this region.  The cyclone B1 near $50-55^\circ$W shows an interior spiral pattern, and existed immediately next to an anticyclone near $45^\circ$W .  This cyclone-anticyclone pair had been present since at least October 2015 \citep{17fletcher_neb} and marked the westward edge of the NEB expansion region.  }
\label{feb2016}
\end{centering}
\end{figure*}

\begin{figure*}
\begin{centering}
\includegraphics[angle=0,width=1.0\textwidth]{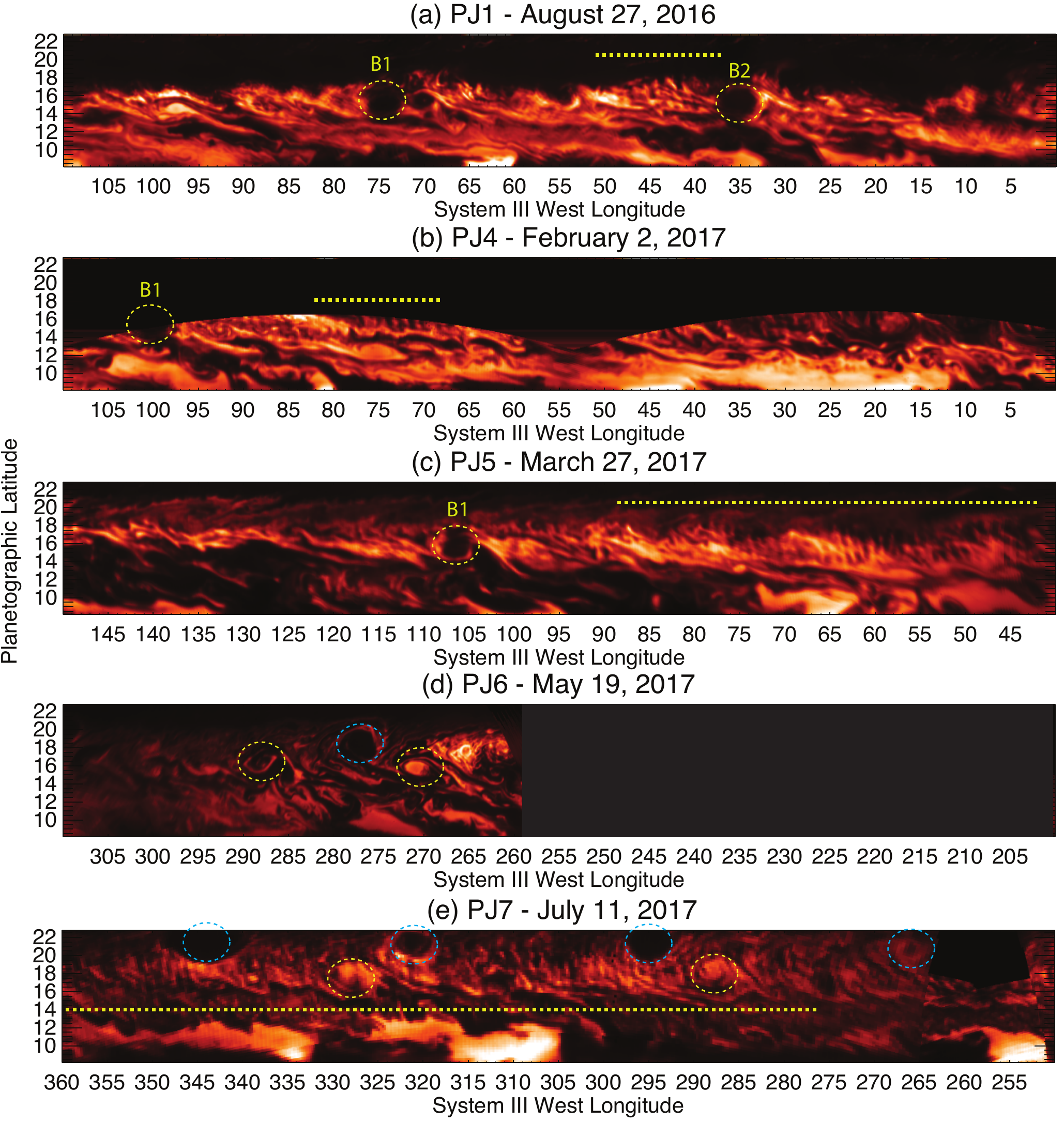}
\caption{Cylindrical projections of Juno JIRAM 5-$\mu$m data from PJ1 to PJ7 (excluding PJ2-3 when JIRAM was not on).  Prominent cyclones are indicated by yellow ovals (B1 and B2 are labelled), prominent anticyclones by blue ovals.  The longitude range of potential mesoscale wave activity is shown by the yellow dotted line, and is most clear during PJ5 data (March 2017) and PJ7 data (July 2017).}
\label{jiram}
\end{centering}
\end{figure*}

\begin{figure*}
\begin{centering}
\includegraphics[angle=0,height=0.9\textheight]{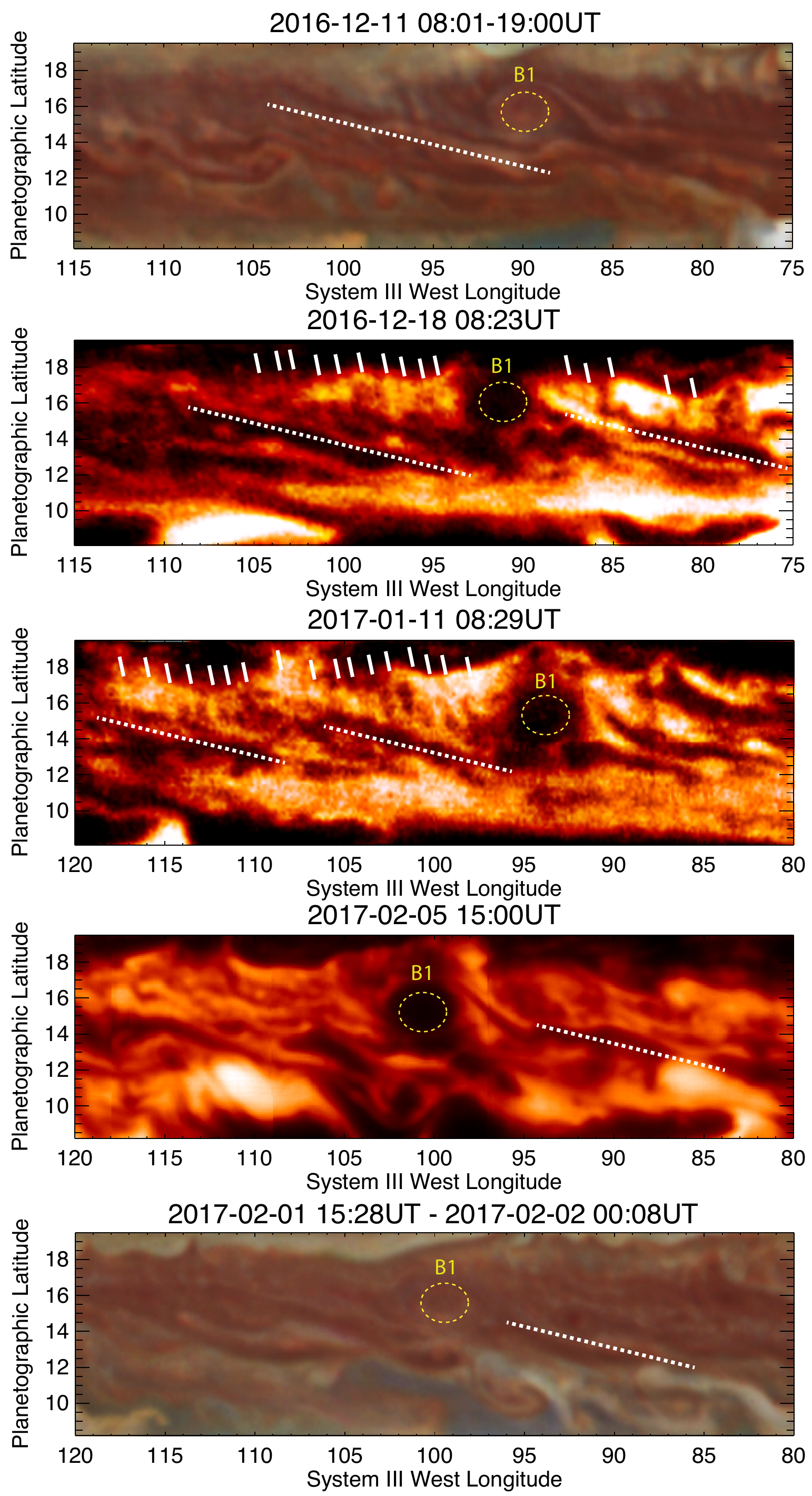}
\caption{Observations of the mesoscale waves over a short longitude segment in December 2016 and February 2017, compared to a Hubble Space Telescope observations on December 11th 2016 and February 1st 2017 (GO-14661).  The NEB was not expanded at this time.  Wave peaks are indicated by solid white lines; cyclone B1 by a yellow dotted oval; and the bright cloud streak (dark at 5 $\mu$m) is indicated by the diagonal white dotted line.  Both HST and 5-$\mu$m images have been sharpened by means of a high-pass filter.  Gemini/NIRI and HST observations in early February show that the wave pattern that had been present in January was no longer visible near B1.}
\label{decjan}
\end{centering}
\end{figure*}

\begin{figure*}
\begin{centering}
\includegraphics[angle=0,height=0.9\textheight]{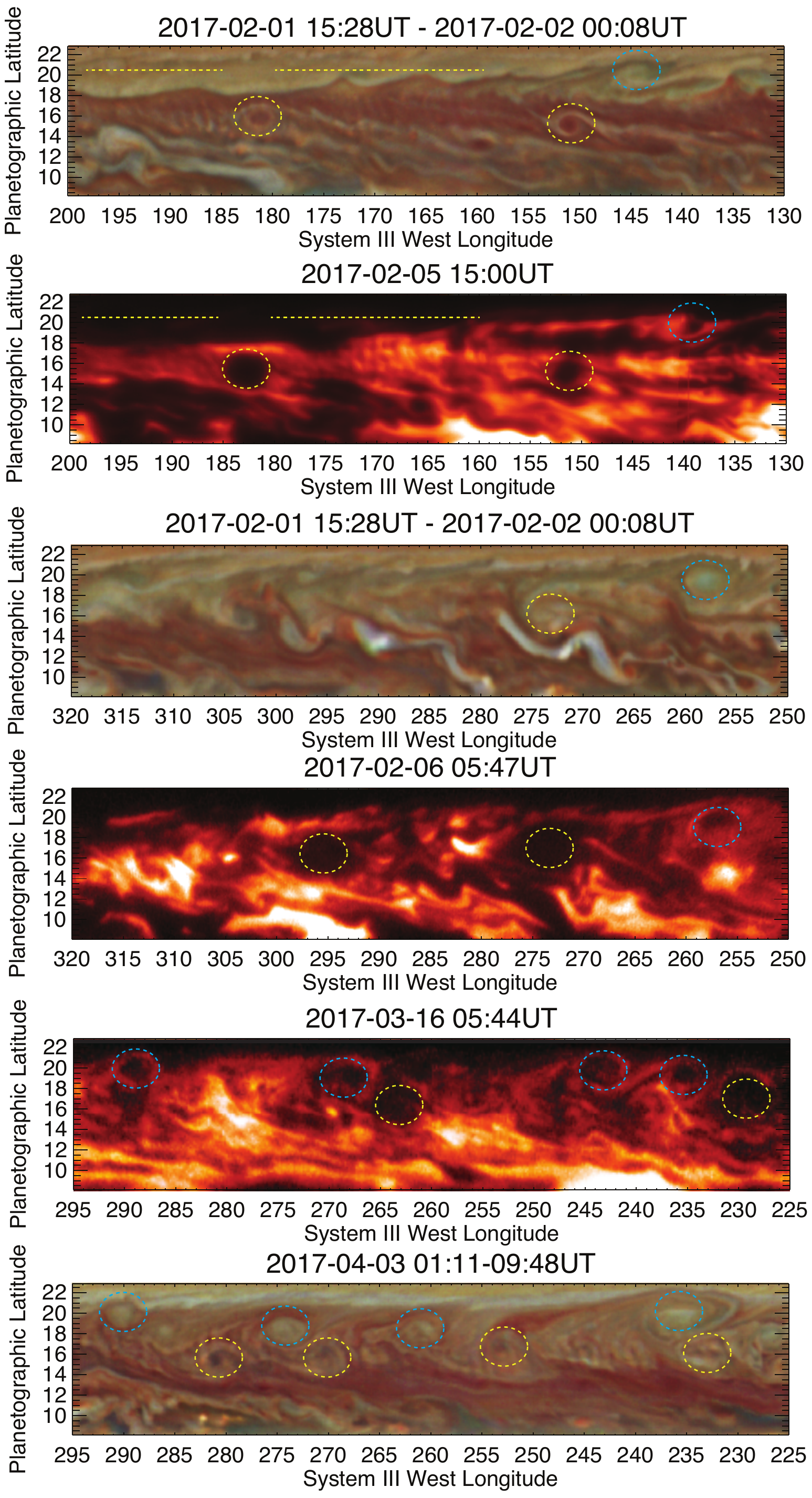}
\caption{VLT and Gemini observations obtained in February and March 2017, once the cyclone-anticyclone pattern was forming.  Waves were still visible in early February at more quiescent longitudes (horizontal yellow dashed line in top panels), but were harder to distinguish at 5 $\mu$m near locations of strong rifting (middle panels) of closely-packed cyclones (yellow ovals) and anticyclones (blue ovals) in the bottom panels.  However, this did not mean that they were absent, as clearly indicated in the HST imaging on April 3rd, 2017 (GO-14756).  Hubble imaging from February 1st 2017 comes from GO-14661.  Both HST and VLT images have been sharpened by means of a high-pass filter.}
\label{febmar}
\end{centering}
\end{figure*}

In this section, we present the chronology of the NEB wave activity as observed at 5 $\mu$m. The prerequisite for observing the waves at this wavelength is the presence of a relatively quiescent background 5 $\mu$m emission that can be modulated solely by the waves, rather than other phenomena like vortices, convective plumes, and `rifts' which appear dark and cloudy against the bright background emission.  For ground-based observers, these conditions were met in December 2016 and January-February 2017 (Fig. \ref{decjan}), when the mesoscale waves were first identified at 5 $\mu$m.  The February 2016 5-$\mu$m observations (Fig. \ref{feb2016}) did not show the presence of waves, whereas the first JIRAM map (August 2016, Fig. \ref{jiram}a) reveals hints of the wave pattern over a limited $12^\circ$ longitude range to the west of a dark cyclone at $35^\circ$W, and these were clearly visible by the end of 2016 from the ground.  

\textit{Timeline of the NEB cyclones: }  Given the potential importance of cyclogenesis in the origins of the wave pattern \citep{15simon}, we now discuss their chronology in detail.  In the August and December images, the waves appear to be located to the west of two cyclones.  Fig. \ref{decjan} shows that the cyclone B1 moved $4^\circ$ westward (from $90^\circ$W to $94^\circ$W) over 30 days between the December and January observations ($0.13^\circ$/day).  This slow motion is consistent with the weak winds in the centre of the NEB.   Extrapolating backwards, this same cyclone B1 is observed at $75^\circ$W in August 2016 (Fig. \ref{jiram}a) and at $55^\circ$W in February 2016 (Fig. \ref{feb2016}).  Indeed, B1 is the same cyclone that marked the westward extension of the stalled 2015/16 expansion, and had been apparent in ground-based data since at least October 2015 \citep{17fletcher_neb}.  This cyclone showed no wave pattern in August 2016, but one had developed by December 2016, 4 months later.  The longevity of cyclone B1, which existed long before waves were identified, suggests that the waves may not originate from cyclone formation, but rather from processes at work once the cyclone was already mature.  B1 was last definitively observed at $106^\circ$W in March 2017 (Fig. \ref{jiram}c), at which time its wave train was no longer visible (it is visible in February 2017 in Gemini/NIRI observations, Fig. \ref{decjan}).   The second cyclone B2 ($35^\circ$W in August 2016) formed within the previously-expanded sector of the NEB, and was the first to exhibit the wave train in the Juno observations at $35^\circ$W.  It was observed again near $58^\circ$W in December 2016, this time without visible waves, but B2 is not visible in either the January VISIR maps or February 2017 Gemini/NIRI maps (Fig. \ref{cmaps_north}), suggesting that the cyclone had dissipated.  

\textit{Early views of the waves:  }  Observations between August 2016 (JIRAM), December 2016 and January 2017 (VLT) suggest that the waves occurred in longitudinally-confined packets. The NEB had its regular width ($7-17^\circ$N) at this stage, before the onset of new 2017 expansion event.  On August 27th, 2016, the waves extended some $12^\circ$ west of cyclone B2, but these had vanished by December 2016 when B2 was at $58^\circ$W, indicating a lifetime shorter than 4 months.  Conversely, the December data showed a longitudinally-confined wave train between $80-105^\circ$W that had not been present in August.  Hubble imaging on December 11th (Fig. \ref{decjan}) also reported the wave between $70-100^\circ$W \citep{18simon}.  The waves appeared to exist within two $10^\circ$-longitude-wide packets, on either side of cyclone B1 at $92^\circ$W, that were latitudinally wider to the east ($\sim3^\circ$) than to the west ($\sim1^\circ$).  This tapering could be related to a dark striation $1^\circ$ further south in Fig. \ref{decjan}, which extended from the southeast to the northwest.  The wavelength was $1.3^\circ$ in the January image, and exhibited a slight tilt from southeast to northwest.    The cyclone B1 and wave pattern moved $\sim2^\circ$ westward (to $94^\circ$W) by January 11th.   The waves retained their $1.2-1.4^\circ$ wavelength, northwesterly tilt, and extended from $95-117^\circ$W longitude.  The latitudinal extent remained $\sim2^\circ$ across the wave train.  Once again, there is a qualitative suggestion that they existed in two distinct packets of approximately $10^\circ$ longitude, although this could be a property of the underlying background emission rather than of the wave itself.  JIRAM observations on February 2nd, 2017 (PJ4) only caught a small glimpse of the wave between $68$ and $82^\circ$W, extending over $\sim14^\circ$ longitude (Fig. \ref{jiram}b), and they were also detectable in Gemini/NIRI observations on February 5th.  There is a considerable amount of small structure at the same spatial scale that could support the presence of a more extensive wave pattern, as suggested by Hubble imaging on February 1st, which reports waves from $25$ to $75^\circ$W \citep{18simon}.  

\textit{Mature wave pattern:}  In February-March 2017 the NEB had expanded northwards to $21^\circ$N (Fig. \ref{cmaps_north}, \ref{febmar}), and a cyclone-anticyclone pattern formed from the `bulges' on the northern edge of the NEB that characterise expansion phases\footnote{We note that the cyclone-anticyclone pattern has the same $22-24^\circ$ wavelength as the strong upper tropospheric thermal wave that was present in 2016, and which was anticorrelated with reflectivity from the upper tropospheric hazes \citep{17fletcher_neb}. }  \citep{17fletcher_neb}. The February images reveal dark cyclones every $\sim22^\circ$ of longitude in the $250-340^\circ$ range at $16^\circ$N (Fig. \ref{febmar}).  By March, these had been joined by anticyclones near $20^\circ$N, with the anticyclones generally occurring to the northwest of the cyclones, although this pairing was not always consistent.  The presence of this cyclone-anticyclone pattern caused the surrounding NEB to become turbulent and chaotic.  With these features strongly modulating the background 5-$\mu$m flux, it proved impossible to observe the mesoscale waves in the $220-320^\circ$W longitude range sampled by the VLT data.  

So was the fine-scale ($1.2^\circ$ wavelength) wave pattern truly absent when the cyclone-anticyclone wave pattern ($22-24^\circ$ wavelength) developed, or is this just a consequence of the visibility at 5 $\mu$m?  Hubble imaging on February 1st and Gemini/NIRI imaging on February 5th showed waves in a relatively quiescent sector of the NEB over $160-200^\circ$W (Fig. \ref{febmar}, top) but not over the more complex $250-320^\circ$W domain sampled by VLT on February 6th (Fig. \ref{febmar}, middle).  This is consistent with turbulent activity removing all signatures of the waves.  Further Hubble imaging on April 3rd showed waves at (a) $5-55^\circ$W and (b) $235-305^\circ$W (Fig. \ref{febmar}, bottom).  The former (a) are readily visible in the March 27th JIRAM observations (Fig. \ref{jiram}c), spanning $45-85^\circ$W.  But the latter (b) were not visible in the VLT images 18 days earlier (Fig. \ref{febmar}).  This suggests that the waves can still be present, but not visible at 5 $\mu$m due to strong variation in the underlying emission.   The Hubble observations in April 2017 \citep{18simon} show a large-scale NEB rifting event (turbulent white cloud structures, sheared east near the NEBs and west near the NEBn) that spanned from $90-220^\circ$W during this period, and it is notable that the waves are only visible outside of this longitude range.  

Hubble imaging and amateur observations \citep{18simon}, as well as Juno observations from PJ7 (Fig. \ref{jiram}e), suggest that the NEB wave was present away from the prominent rifting zone until at least July 2017.  Waves were not visible in the $260-305^\circ$W range sampled by JIRAM during PJ6, but were visible 7-8 weeks later spanning from $280^\circ$W to $30^\circ$W, confirming that they can develop and disappear over monthly timescales.  We can conclude that the mesoscale waves form only in a relatively quiescent NEB, away from prominent rifts, and initially in association with pre-existing cyclones.

%\textit{Generate a cartoon showing the wave longitudes as a function of time}.

\section{Discussion}
\label{discuss}

\subsection{Detection at 5 $\mu$m}
The primary conclusion of this work is that the mesoscale wave activity observed at visible wavelengths \citep{15simon, 18simon} also serves to modulate the 5-$\mu$m emission from Jupiter's deeper troposphere.  This provides a new window for tracking this unusual phenomenon.  However, the detection alone need not necessarily imply that the wave pattern is deep.  Firstly, there is a non-negligible contribution from reflected sunlight at 5 $\mu$m, which reflects off of the same tropospheric clouds that are evident in the visible-light imaging \citep{15bjoraker}.  A modulation of the upper tropospheric clouds could therefore influence the reflected sunlight contribution, but aerosols within the belts are expected to have a low albedo at 5 $\mu$m.  However, the upper tropospheric aerosols can modulate the 5-$\mu$m radiance in the absence of reflected sunlight, simply via absorption and scattering.  \citet{15giles} required a cloud at  $p<1.2$ bar to reproduce Cassini VIMS spectra of both cloudy zones and cloud-free belts.  Lower pressures were also permitted by the data, implying that this cloud could reside at the NH$_3$-condensation level near 800 mbar.  Furthermore, Galileo NIMS analyses by \citet{01irwin, 01nixon, 02irwin} place the primary cloud decks in the 1-2 bar range; and \citet{10sromovsky} suggested a spatially-variable cloud base between 0.79-1.27 bar.   Although optical thickness variations of a deeper cloud were also required to fit the 5-$\mu$m spectra, we have no way to determine whether the mesoscale wave modulation exists in the upper layer, the deeper layer, or both.  

Each of these works are consistent with a possible source of 5-$\mu$m modulation in the 0.8-1.2 bar range, without having to assume that the mesoscale wave is present at higher pressures.  Near-infrared spectroscopy of the NEB is required to better constrain the altitude of the wave pattern, as described by \citet{18adriani} using JIRAM data.  Their analysis also suggests that the data can be explained by waves in the upper layer, the deeper layer, or a combination of both.   Although we cannot distinguish these possibilities, we favour a lower pressure for the reasons outlined above.

Finally, the calibrated Juno/JIRAM datasets allow us to provide a quantitative estimate of the brightness amplitude of the wave pattern.  The May 2017 observations (PJ5) have a mean brightness temperature at $16^\circ$N of $210\pm15$K, where the large range is a result of the extreme variability seen in Fig. \ref{jiram}c.  The wave amplitude is 7-10 K peak to peak (approximately 2 $\mu$W/cm$^2$/sr/$\mu$m), where we caution the reader that this is a brightness temperature resulting from aerosol modulation, rather than a physical temperature variation.  The cyclone near $106^\circ$W reaches brightness temperatures of $\sim195$ K compared to a background of $\sim220$ K.  We find the same range of brightness temperatures in the PJ7 data.  Via an analysis of Cassini/VIMS spectra, \citet{15giles} demonstrated that changes of 2 $\mu$W/cm$^2$/sr/$\mu$m could easily be reproduced by changes in the opacity of a $p=0.8$ bar cloud of 10-20\%, which suggests that upper tropospheric modulation of the deeper 5-$\mu$m radiance is a reasonable hypothesis.

\subsection{Wave origins and nature}
The chronology of the NEB wave described above raises several intriguing possibilities, which we summarise here.  

\begin{figure*}
\begin{centering}
\includegraphics[angle=0,width=1.1\textwidth]{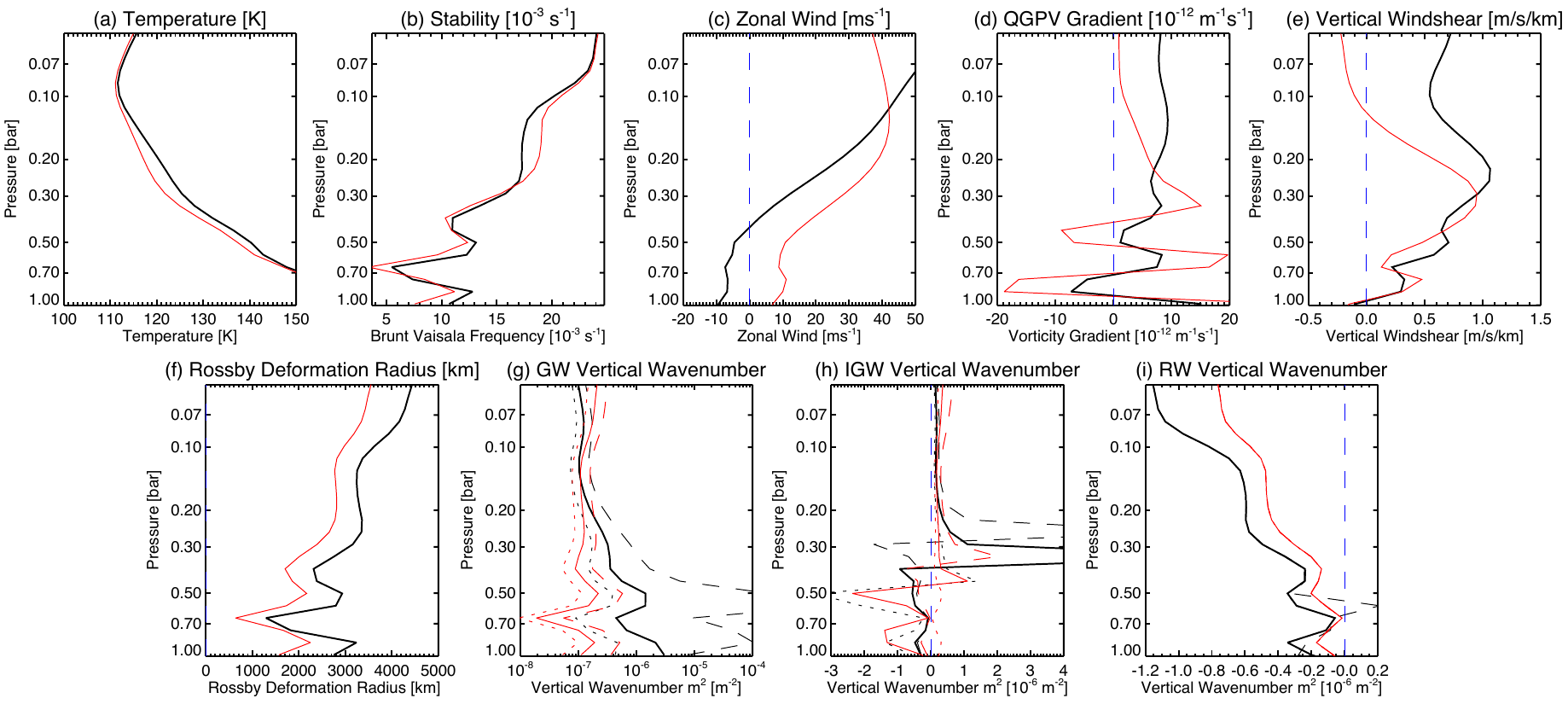}
\caption{Exploring the properties of the NEB using Cassini/CIRS temperature retrievals from 2000 \citep{16fletcher_texes}, comparing the latitude of the NEB cyclones ($16^\circ$N, solid black line) and that of the NTrZ anticyclones ($20^\circ$N, solid red line).  We plot (a) the temperatures; (b) the Brunt V\"{a}is\"{a}l\"{a} frequency; (c) the zonal winds estimated from the vertical windshear in (e); (d) the latitudinal gradient of the quasi-geostrophic potential vorticity; (f) the Rossby Deformation Radius; and the square of the vertical wavenumber $m$ in the case of pure gravity waves (g), inertio-gravity waves (h) and Rossby waves (i). CIRS constrained temperatures between 80-800 mbar, so values outside of this range should be treated with caution.  For the calculations of $m^2$, we show estimates for $c_x=-5$ m/s (dashed line), $c_x=-15$ m/s (solid line) and $c_x=-25$ m/s (dotted line) for $16^\circ$N (black) and $20^\circ$N (red).  Parameters based on first- and second-order derivatives of the $T(p)$ are highly uncertain, as described in the main text, and should be considered as a qualitative guide only.  The blue dashed line indicates the zero point.} 
\label{NEB_dyn}
\end{centering}
\end{figure*}

\begin{enumerate}

\item Mesoscale waves were first observed at 5 $\mu$m in mid-2016 in association with two cyclones, B1 and B2, that were related to the stalled 2015-16 expansion of the NEB.  B1 had been present since at least October 2015, and had been part of a cyclone-anticyclone pair that marked the western edge of the expanded NEB sector.  Given the rarity of the mesoscale waves, and their re-detection during a period when the NEB has been observed to sporadically expand into the NTrZ, we speculate that the NEB expansion conditions (and the resulting cyclogenesis) may be influencing the environmental conditions permitting the propagation of these waves.

\item The wave trains were initially restricted in longitude to packets $\sim10^\circ$ wide, meaning that only $\sim7-10$ wave crests were visible in the early detections.  These packets appeared to the west of B1 and B2, and had lifetimes of a few months.  The packets were latitudinally wider in the east than in the west, and the wave crests appeared to be sheared from southeast to northwest.  Such a morphology is likely related to the changing wind across the NEB (centred on the westward NEBn jet at $17^\circ$N), and also to the visibly-bright and 5-$\mu$m dark elongated rifts that could be seen in Fig. \ref{decjan} stretching across the NEB from southeast to northwest.  The wave train expanded to span $\sim40-50^\circ$ longitude some 6-8 months after they were first detected, and were most visible in March and April 2017.  

\item The waves were not visible when the NEB became chaotic.  At 5-$\mu$m, the presence of a closely-packed pattern of cyclones and anticyclones (typical of the end stages of an NEB expansion and contraction event) rendered the small-scale wave pattern invisible, although it could still be detected in visible light.  However, strong rifting activity that is bright in the visible and dark at 5 $\mu$m (which was present from $90-220^\circ$W in March 2017) caused the wave pattern to be invisible at all wavelengths, and potentially disrupted it completely.  We suggest that relatively quiescent NEB conditions are therefore needed for the wave pattern to propagate.

\item The waves retained their $1.1-1.4^\circ$ wavelength (a wavenumber of 260-330) at all epochs, without notable increases or decreases with time.  \citet{15simon} point out that this, and the $\sim2^\circ$ latitudinal extent of the waves, are on the same scale as the atmospheric deformation radius (see Fig. \ref{NEB_dyn}f), which may be playing a role in setting the length scale for these waves. 

\end{enumerate}

By analogy to waves associated with cyclones and anticyclones on Earth and in General Circulation Models, \citet{18simon} propose both pure gravity waves (GWs, under the action of gravity and buoyancy forces) and/or inertio-gravity waves (IGWs, with wavelengths long enough to require consideration of the Coriolis term) as plausible explanations for the observed mesoscale wave pattern. In this analysis we also consider short-wavelength Rossby waves (RWs), although we note that jovian Rossby waves typically have much longer wavelengths and a very different morphology to the mesoscale waves described here \citep[e.g., NEB thermal waves with a $22-24^\circ$ longitude wavelength were identified in the NEB in 2015-16,][]{17fletcher_neb}.  We utilise Jupiter's zonal-mean temperature structure retrieved from Cassini Composite Infrared Spectrometer (CIRS) data by \citet{16fletcher_texes} to explore the dispersion relationships of these three wave types.  Although more recent Earth-based temperature observations are available, their analysis is ongoing and the Cassini thermal structure from December 2000 can be used as a good qualitative proxy for present-day conditions within the NEB.  

%For a barotropic Rossby wave in the zonal (east-west) direction \citep{11sanchez}, $u-c_x=\beta/k^2$, where $\beta$ represents the change in the Coriolis force with latitude, $c_x$ is the longitudinal phase speed of the waves \citep[assumed to be -3.7 m/s, westward,][]{18simon}; $u$ is the zonal wind determined from the thermal windshear $dT/dy$ (where $y$ is the north-south distance); and $k=2\pi/L_x$ is the zonal wavenumber.  

We take the dispersion relationship for linear GWs as \citep{00holton, 11sanchez}: 
\begin{equation}
\omega^2= (c_x-u)^2k^2 = \frac{N^2k^2}{k^2+m^2+1/(4H^2)}
\end{equation}
Here $\omega$ is the intrinsic frequency; $c_x$ is the longitudinal phase speed of the waves; $u$ is the zonal wind determined from the thermal windshear $dT/dy$ (where $y$ is the north-south distance); $k$ and $m$ are the zonal and vertical wavenumbers, respectively; $N$ is the Brunt V\"{a}is\"{a}l\"{a} frequency; and $H$ is the scale height.  Note that we take the meridional wavenumber to be zero, as the wave crests are nearly zonal, i. e. perpendicular to longitude and flow direction. The dispersion relationship for IGWs (e.g., accounting for the Coriolis parameter $f$) is given by \citep[Section 4.6.3, ][]{87andrews}:
\begin{equation}
%\omega^2= (c_x-u)^2k^2 = f^2 + \frac{N^2(k^2+l^2)}{m^2+1/(4H^2)}
\omega^2= (c_x-u)^2k^2 = f^2 + \frac{N^2k^2}{m^2+1/(4H^2)}
\label{m2IGW}
\end{equation}
Finally, the dispersion relationship for a quasi-geostrophic Rossby wave (RW) reduces to \citep{11sanchez}:
\begin{equation}
c_x-u = - \frac{\beta_e}{k^2+(f^2/N^2)(m^2+(1/4H^2))}
\end{equation}
Here we have introduced the latitudinal gradient of the quasi-geostrophic potential vorticity, \citep[$\beta_e=dq_G/dy$, the `effective beta',][]{87andrews}, evaluated followed the method of \citep{16fletcher}.  This is related to the change in the Coriolis parameter with latitude, which provides the restoring force for Rossby waves.

Note that we consider only GWs, IGWs and RWs, rather than Kelvin-Helmholtz instabilities, because the Richardson number ($N^2/(du/dz)^2$) is positive and much larger than one throughout the 80-800 mbar range sampled here.   These dispersion relationships above rely on parameters that can be derived from the CIRS temperature measurements in the 0.1-0.8 bar range:  zonal winds are estimated from the thermal wind equation \citep{87andrews}, and the buoyancy frequency utilised the measured and dry adiabatic lapse rates (using the local gravity and specific heat capacity).  The mean zonal phase speed of the waves, $c_x$ is estimated to be $-15.5\pm10$ m/s \citep[from an average of the 2016-17 measurements of][]{18simon}, implying a slow westward motion with respect to the cloud-tracked zonal winds at $16^\circ$N ($u\sim-10$ m/s in the centre of the NEB, Fig. \ref{NEB_dyn}c).  These dispersion relationships should strictly rely on constant $N$ and $u$ with altitude, which is not the case in Fig. \ref{NEB_dyn}, so should be considered only as approximations.  These equations can be rearranged to estimate $m$ to assess the likelihood of vertical propagation, first for GWs:
\begin{equation}
m^2=\frac{N^2}{(c_x-u)^2}-k^2-\frac{1}{4H^2}
\end{equation}
secondly for IGWs:
\begin{equation}
%m^2=\frac{N^2(k^2+l^2)}{(c_x-u)^2k^2 - f^2} - \frac{1}{4H^2}
m^2=\frac{N^2k^2}{(c_x-u)^2k^2 - f^2} - \frac{1}{4H^2}
\end{equation}
and finally for RWs:
\begin{equation}
m^2=\frac{N^2}{f^2}\left(\frac{\beta_e}{u-c_x}-k^2\right)-\frac{1}{4H^2}
\label{m2RW}
\end{equation}
If $m^2>0$ then real solutions can be found, and vertical propagation of this wavetype is permitted.  If $m^2<0$ then the solutions are imaginary, and the wave could be trapped in the vertical at the level that it is observed.  

Fig. \ref{NEB_dyn} shows these calculated quantities - temperatures, windshears, Rossby Deformation Radius ($L_D=NH/f$) and $m^2$ - for the GW, IGW, and RW cases.  These are provided at $16^\circ$N (the latitude of the cyclone chain in the NEB) and $20^\circ$N (the latitude of the associated anticyclones in the NTrZ).  Before describing the results, we first caution the reader that the parameters in Fig. \ref{NEB_dyn} are subject to large uncertainties - windshears and static stability both require derivatives of the retrieved temperatures with latitude and altitude, and estimates of the winds (and associated vorticity gradients) require integration of these gradients with altitude.  This can serve the magnify the 1-2 K uncertainties on the $T(p)$ profile in Fig. \ref{NEB_dyn}a, even if we considered the CIRS inversions to be `ideal' \citep{16fletcher}.  Additional errors arise from assumptions on vertical smoothing and variable information content with height, zonal averaging over Jupiter's spatially-variable temperatures \citep{16fletcher_texes}, and the altitude level of the cloud-tracked zonal winds.  Quantifying the effects of these unknowns in a meaningful way is challenging, but the parameters in Fig. \ref{NEB_dyn} represent our best estimate from the CIRS data, and are sufficient to show qualitative trends, rather than quantitative values. 

With these caveats in mind, we explore the implications of the vertical wavenumbers in Fig. \ref{NEB_dyn}.  In the GW case, we find that vertical propagation is allowed at all altitudes sampled ($m$ is always real in Fig. \ref{NEB_dyn}g).  The value of $m$ varies considerably within the $c_x=-(15\pm10)$ m/s uncertainly envelope quoted by \citet{18simon}, but the central value of $m\sim1\times10^{-3}$ m$^{-1}$ near 500-700 mbar is equivalent to a vertical wavelength of $\sim6$ km, or a third of a scale height.  For IGWs in Fig. \ref{NEB_dyn}h, we find that $m^2<0$ for high pressures, but that vertical propagation is allowed for lower pressures.  The transition between these regimes is sensitive to $c_x$, with more westward phase speeds ($c_x\approx-25$ m/s) able to propagate vertically at deeper pressures ($p<500$ mbar), whereas slower phase speeds ($c_x\approx-5$ m/s) could only propagate in the upper troposphere ($p<300$ mbar).  Closer inspection of Equation \ref{m2IGW} shows IGW vertical propagation is inhibited when the intrinsic frequency $\omega$ is smaller than the Coriolis parameter $f$, which is the case for $p>400$ mbar for waves with a $c_x=-15$ m/s phase speed.  IGWs are not conclusively ruled out for higher pressures ($p>500$ mbar), provided they form at the altitude at which they are observed, which is estimated to be $\sim500$ mbar from Hubble data \citep{18simon}.  Finally, the vertical wavenumber for RWs is found to be negative almost everywhere - this is because $k^2>>\beta_e/(u-c_x)$ in Equation \ref{m2RW} for this short-period wave (so the value of $c_x$ has negligible influence on $m^2$), with the second term only gaining prominence when $u\approx c_x$, which is the case for $c_x=-5$ m/s in Fig. \ref{NEB_dyn}i.  Given that the mesoscale wave bears little resemblance to Rossby waves previously identified on Jupiter, and the fact that vertical propagation is prohibited, we deem this to be the least likely explanation for these waves.  Given that both $c_x$ and the background zonal $u$ from the thermal wind equation are significantly uncertain, these estimates of $m^2$ should be considered as qualitative guides only.

%

%\textbf{Although this quantity has significant uncertainties, }this is consistent with a vertically-confined GW pattern in the upper tropospheric aerosol layers (400-500 mbar), modulating both the Hubble reflectivity and the 5-$\mu$m transmission.     However, the magnitude of $m^2$, and the range of $m^2<0$ in the IGW case, are sensitive to the estimate of $c_x$ and $u$, and vary significantly within the $\pm10$ m/s uncertainty quoted by \citet{18simon}.  Having $c_x-u$ in the denominator of both equations can lead to singularities in the calculation when $c_x\approx u$, which is the case in Fig. \ref{NEB_dyn}c near 400-500 mbar.    The quantitative estimates given above should really be considered only as qualitative guides.

Frustratingly, further progress on identifying the nature of this wave pattern cannot be extracted from these measurements alone.  Preliminary GCM modelling \citep{18simon} is similarly inconclusive.  Both GWs and IGWs are plausible explanations, with the former propagating more easily in the vertical throughout the upper troposphere.  It is reasonable to hypothesise that the 5-$\mu$m brightness and the Hubble reflectivity are being modulated by \textit{the same aerosol layers} somewhere in the 400-800 mbar range.  Finally, it has been previously noted that the latitudinal gradient of quasi-geostropic potential vorticity \citep[$\beta_e=dq_G/dy$, the `effective beta',][]{87andrews} changes sign across the NEB \citep{06li, 16rogers_wave}.  Fig. \ref{NEB_dyn}d shows a complex vertical structure at $16-21^\circ$N, with the gradient changing sign several times in the 400-1000 mbar range.  Such sign changes are a necessary (but not sufficient) condition for baroclinic instability following the Charney-Stern criterion \citep{62charney}.   Whilst not conclusive, it is at the very least suggestive of mesoscale GWs or IGWs related to instabilities in the upper tropospheric aerosol layers (400-800 mbar), modulating both the Hubble reflectivity and the 5-$\mu$m brightness.

%the CIRS-derived thermal structure offers some insights into the latitudinal location of this wave.  Evaluating $m^2$ in the GW case indicates that conditions for vertical propagation of a wave of this scale ($k$ and $c_x$) is only plausible between $\sim15-19^\circ$N, although the calculation is extremely sensitive to noise.  This wave would be unlikely to propagate vertically if located at $20^\circ$N (dotted line).  Although the argument is circular, it is at least consistent, with a mesoscale wave of this type being present at $16^\circ$N.  Furthermore,

\section{Conclusions}

Using a technique of `lucky imaging' to freeze atmospheric seeing, the VLT/VISIR and Gemini/NIRI instruments have provided the highest-resolution 5-$\mu$m views of Jupiter's atmosphere obtained from Earth.  Coupled with Juno/JIRAM M-band maps from the first seven Juno perijoves (August 2016-July 2017), and Hubble and amateur imaging at visible wavelengths \citep{18simon}, this has enabled the detection of a mesoscale wave pattern in Jupiter's North Equatorial Belt, often (but not always) associated with a chain of cyclones at $16^\circ$N (within the NEB) and associated anticyclones at $19-20^\circ$N (in the NTrZ).  The genesis and evolution of this wave pattern is unclear - the waves have a longitudinal wavelength of $1.1-1.4^\circ$ (wavenumber 260-330, corresponding to a length of $\sim$1,300-1,600 km) that does not change with time; the wave crests are aligned north-south with a slight north-westward tilt; they cause a 7-10 K brightness temperature modulation at 5 $\mu$m consistent with 10-20\% changes in the opacity of an upper tropospheric cloud; they exist over a limited $\sim2^\circ$ latitude range near the northern edge of the NEB; and they start small (initially detected in longitudinal packets extending $\sim10^\circ$ west of cyclones) but evolve to span a broad range of longitudes.  The waves appear to be ephemeral, appearing and disappearing on timescales of weeks.  They are removed or rendered invisible by chaotic rifting activity within the NEB, which is present in March 2017.  

The true nature of the waves remain elusive.  The thermal structure of the NEB was used to investigate gravity wave (GW), inertio-gravity wave (IGW), and Rossby wave (RW) dispersion relationships, finding that GWs are able to propagate vertically throughout the upper troposphere, whereas IGW propagation is only permitted at low pressures, and RW propagation is ruled out throughout this domain for a wave of this small scale.  We cannot definitively rule out a wave source at higher pressures, where information on the thermal structure (and wave propagation conditions) is unavailable.  But it is plausible that mesoscale waves (GWs or IGWs) could be modulating Jupiter's upper tropospheric aerosols in the 400-800 mbar range.  These aerosols are detected through their reflectivity \citep{18simon} and their attenuation of 5-$\mu$m radiance originating from deeper atmospheric pressures.  We note that this region of the NEB exhibits the necessary conditions to violate a range of instability criteria, favouring wave genesis at this latitude.  

The linkage between these waves and the NEB cyclones is compelling but inconclusive, pending future numerical simulations.  The presence of the waves today, compared to their rarity in previous years, could be related to the recent 2015-16 and 2016-17 expansion and contraction episodes of the North Equatorial Belt \citep{17fletcher_neb}, and we note that the waves were first spotted in association with two cyclones that had played a prominent role in the 2015-16 expansion.   Changes to the tropospheric thermal conditions and their correlation with the presence of the mesoscale waves could provide insights into their genesis, and will be the subject of future intensive study.

%%%%%%%%%%%%%%%%%%%%%%%%%%%%%%%%
%%%%%%%%%%%%%%%%%%%%%%%%%%%%%%%%
%%%%%%%%%%%%%%%%%%%%%%%%%%%%%%%%
\acknowledgments
Fletcher was supported by a Royal Society Research Fellowship and European Research Council Consolidator Grant (under the European Union's Horizon 2020 research and innovation programme, grant agreement No 723890) at the University of Leicester.  The UK authors acknowledge the support of the Science and Technology Facilities Council (STFC).   Orton was supported by grants from NASA to the Jet Propulsion Laboratory, California Institute of Technology.  A. Sanchez-Lavega and R. Hueso were supported by by the Spanish projects AYA2015-65041-P (MINECO/FEDER, UE) and Grupos Gobierno Vasco IT- 765-13.  We are grateful to R. Morales-Juberias and R. Cosentino for discussions on the contents of this article.  This investigation was based on thermal-infrared observations acquired at  the ESO Very Large Telescope Paranal UT3/Melipal Observatory, with program IDs 60.A-9620, 098.C-0681 and 099.C-0612.    Observations were also obtained at the Gemini Observatory (program ID GN-2017A-Q-60), which is operated by the Association of Universities for Research in Astronomy, Inc., under a cooperative agreement with the NSF on behalf of the Gemini partnership: the National Science Foundation (United States), the National Research Council (Canada), CONICYT (Chile), Ministerio de Ciencia, Tecnología e Innovación Productiva (Argentina), and Ministério da Ciência, Tecnologia e Inovação (Brazil).  The JIRAM project was founded by the Italian Space Agency (ASI), and we are grateful to all those who participated in the design of these data.  VLT observations are available through the ESO archive\footnote{\url{http://archive.eso.org/eso/eso\_archive\_main.html}}, JIRAM observations are available through the Planetary Data System Atmospheres Node\footnote{\url{https://pds-atmospheres.nmsu.edu/data\_and\_services/atmospheres\_data/JUNO/jiram.html}}.  The lucky imaging made use of the Autostakkert software\footnote{\url{autostakkert.com}}.  HST observations are associated with programs GO-14661 and GO-14756, with support provided to Simon, Wong, Barnett, and Orton by NASA through grants from the Space Telescope Science Institute (operated by the Association of Universities for Research in Astronomy, Inc., under NASA contract NAS 5-26555). HST data can be obtained from the MAST archive\footnote{\url{https://archive.stsci.edu/prepds/opal}, \url{https://archive.stsci.edu/prepds/wfcj}}.

\vspace{5mm}
\facilities{VLT, Juno, Hubble, Gemini}


%\software{astropy \citep{2013A&A...558A..33A},  
%          Cloudy \citep{2013RMxAA..49..137F}, 
%          SExtractor \citep{1996A&AS..117..393B}
%          }



\appendix

\section{Status of Jupiter's Major Belts}

\begin{figure*}
\begin{centering}
\includegraphics[angle=0,width=1.0\textwidth]{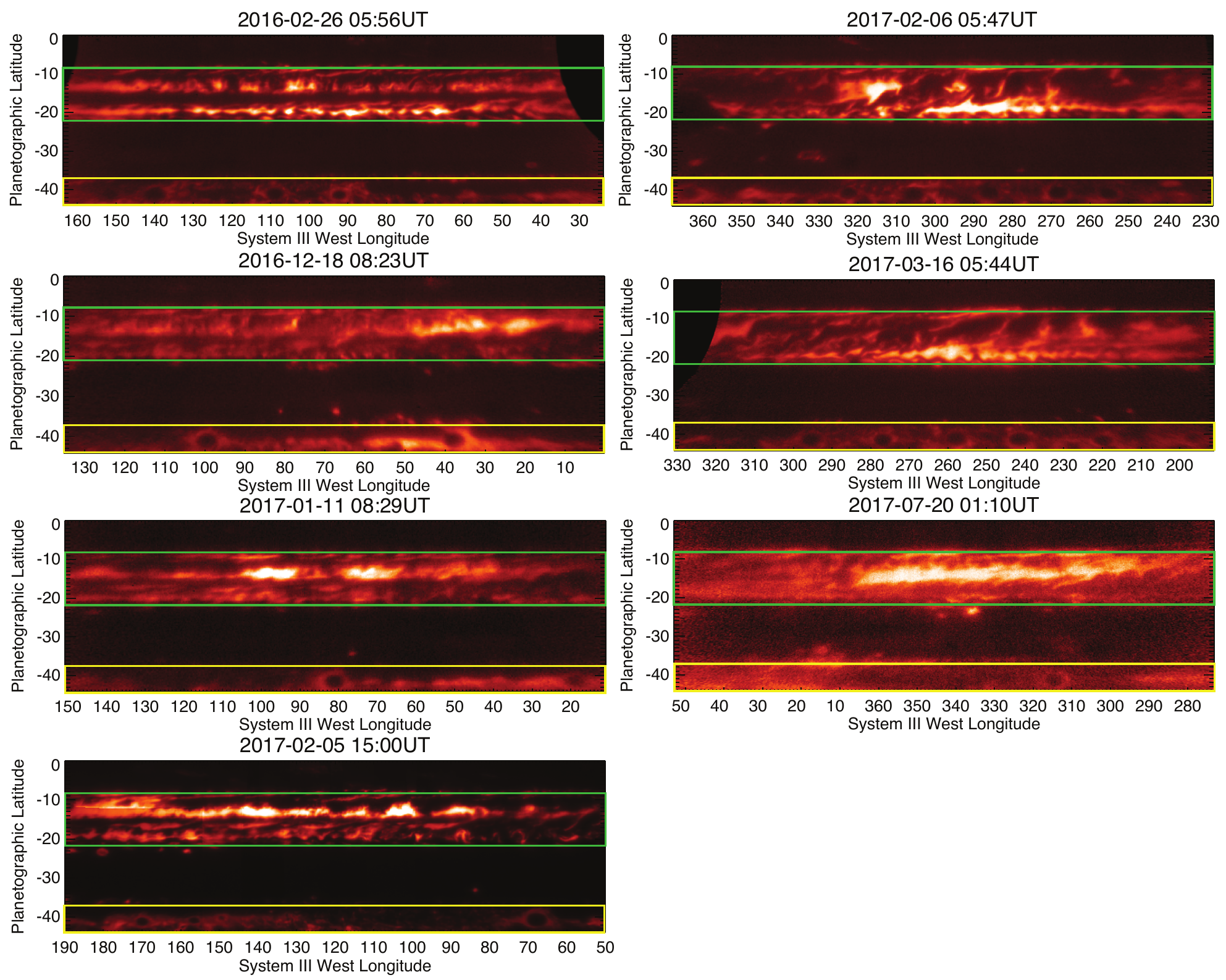}
\caption{Maps of 5-$\mu$m emission in the southern hemisphere, highlighting the South Equatorial Belt (SEB) region in green and the South South Temperate Belt (SSTB) region in yellow.  }
\label{cmaps_south}
\end{centering}
\end{figure*}

To provide context for the changes observed in the NEB, we now describe additional insights available in the 5-$\mu$m maps at other latitude bands:

\begin{itemize}

\item South Equatorial Belt (SEB):  Fig. \ref{cmaps_south} reveals how the SEB emission had a banded appearance in early 2016, but that this was significantly disrupted by an outbreak of convective plumes in early 2017.  Although the SEB was not in a faded state \citep{11fletcher_fade}, the mid-SEB outbreak produced 5-$\mu$m dark clouds and bright peripheral lanes that extended between the prograde SEBn jet and the retrograde SEBs jet in a characteristic `S' shape reminiscent of the 2010-11 revival \citep{17fletcher_seb}.  The outbreak occurred near $16^\circ$S, $300^\circ$W in late December 2016, and dominates the appearance of the SEB in February and March 2017 (region [4] of Fig. \ref{globes}).

\item South South Temperate Belt (SSTB):  A chain of anticyclonic vortices can be seen in the SSTB, appearing white in visible light and dark (i.e., cloudy) at 5 $\mu$m.  They are embedded in a belt of diffuse emission, and each displays peripheral rings associated with subsidence and aerosol clearing \citep[e.g.,][]{10depater_jup}.  These large anticyclonic ovals dominate the temperate southern latitudes and are resolved in good detail by the VISIR burst mode imaging (region [5] of Fig. \ref{globes}).  Note that Oval BA does not exhibit any peripheral ring (it is located near $320^\circ$W in February-March 2017), suggesting that the cloud-coverage of the southern temperate zones is too large to permit 5-$\mu$m emission.  A faint cyclonic region immediately west of Oval BA can be observed in emission near $33^\circ$S, $332^\circ$W in February 2017.

\item North Temperate Belt (NTB):  Region [1] of Fig. \ref{globes} and Fig. \ref{cmaps_north} indicate that plume activity on the southern edge of the NTB in October 2016 \citep{17sanchez_NTB, 17hueso_jup} corresponded to a change in the cloud opacity over the northern edge of the NTB.  February 2016 observations, before the NTB outbreak, indicated small and localised spots of emission, suggesting that much of the NTB was clouded over (i.e., faded).  The 2017 observations all indicate a bright, broken band over the NTB(N).  This sits northwards of the deep red NTB(S) band that formed in the aftermath of the October 2016 outbreak, and represents a clearing (revival) of the NTB(N).

\end{itemize}


\begin{thebibliography}{}
\expandafter\ifx\csname natexlab\endcsname\relax\def\natexlab#1{#1}\fi
\providecommand{\url}[1]{\href{#1}{#1}}
\providecommand{\dodoi}[1]{doi:~\href{http://doi.org/#1}{\nolinkurl{#1}}}
\providecommand{\doeprint}[1]{\href{http://ascl.net/#1}{\nolinkurl{http://ascl.net/#1}}}
\providecommand{\doarXiv}[1]{\href{https://arxiv.org/abs/#1}{\nolinkurl{https://arxiv.org/abs/#1}}}

\bibitem[{{Adriani} {et~al.}(2017){Adriani}, {Filacchione}, {Di Iorio},
  {Turrini}, {Noschese}, {Cicchetti}, {Grassi}, {Mura}, {Sindoni}, {Zambelli},
  {Piccioni}, {Capria}, {Tosi}, {Orosei}, {Dinelli}, {Moriconi}, {Roncon},
  {Lunine}, {Becker}, {Bini}, {Barbis}, {Calamai}, {Pasqui}, {Nencioni},
  {Rossi}, {Lastri}, {Formaro}, \& {Olivieri}}]{17adriani}
{Adriani}, A., {Filacchione}, G., {Di Iorio}, T., {et~al.} 2017, Space Sci.
  Rev., 213, 393, \dodoi{10.1007/s11214-014-0094-y}

\bibitem[{{Adriani} {et~al.}(2018){Adriani}, {Moriconi}, {Altieri}, {Sindoni},
  {Grassi}, {Fletcher}, {Melin}, {Mura}, {Ingersoll}, {Atreya}, {Tosi},
  {Cicchetti}, {Noschese}, {Lunine}, {Orton}, {Plainaki}, {Sordini},
  {Olivieri}, {Bolton}, {Connerney}, \& {Levin}}]{18adriani}
{Adriani}, A., {Moriconi}, M.~L., {Altieri}, F., {et~al.} 2018, in preparation

\bibitem[{{Andrews} {et~al.}(1987){Andrews}, {Holton}, \& {Leovy}}]{87andrews}
{Andrews}, D.~G., {Holton}, J.~R., \& {Leovy}, C.~B. 1987, {Middle atmosphere
  dynamics} (Academic Press, New York)

\bibitem[{{Bjoraker} {et~al.}(2015){Bjoraker}, {Wong}, {de Pater}, \&
  {{\'A}d{\'a}mkovics}}]{15bjoraker}
{Bjoraker}, G.~L., {Wong}, M.~H., {de Pater}, I., \& {{\'A}d{\'a}mkovics}, M.
  2015, {Astrophys. J.}, 810, 122, \dodoi{10.1088/0004-637X/810/2/122}

\bibitem[{{Bolton} {et~al.}(2017){Bolton}, {Adriani}, {Adumitroaie}, {Allison},
  {Anderson}, {Atreya}, {Bloxham}, {Brown}, {Connerney}, {DeJong}, {Folkner},
  {Gautier}, {Grassi}, {Gulkis}, {Guillot}, {Hansen}, {Hubbard}, {Iess},
  {Ingersoll}, {Janssen}, {Jorgensen}, {Kaspi}, {Levin}, {Li}, {Lunine},
  {Miguel}, {Mura}, {Orton}, {Owen}, {Ravine}, {Smith}, {Steffes}, {Stone},
  {Stevenson}, {Thorne}, {Waite}, {Durante}, {Ebert}, {Greathouse}, {Hue},
  {Parisi}, {Szalay}, \& {Wilson}}]{17bolton}
{Bolton}, S.~J., {Adriani}, A., {Adumitroaie}, V., {et~al.} 2017, Science, 356,
  821, \dodoi{10.1126/science.aal2108}

\bibitem[{{Charney} \& {Stern}(1962)}]{62charney}
{Charney}, J.~G., \& {Stern}, M.~E. 1962, Journal of Atmospheric Sciences, 19,
  159, \dodoi{10.1175/1520-0469(1962)019<0159:OTSOIB>2.0.CO;2}

\bibitem[{{de Pater} {et~al.}(2010){de Pater}, {Wong}, {Marcus}, {Luszcz-Cook},
  {{\'A}d{\'a}mkovics}, {Conrad}, {Asay-Davis}, \& {Go}}]{10depater_jup}
{de Pater}, I., {Wong}, M.~H., {Marcus}, P., {et~al.} 2010, {Icarus}, 210, 742,
  \dodoi{10.1016/j.icarus.2010.07.027}

\bibitem[{{Fletcher} {et~al.}(2016{\natexlab{a}}){Fletcher}, {Greathouse},
  {Orton}, {Sinclair}, {Giles}, {Irwin}, \& {Encrenaz}}]{16fletcher_texes}
{Fletcher}, L.~N., {Greathouse}, T.~K., {Orton}, G.~S., {et~al.}
  2016{\natexlab{a}}, Icarus, 278, 128, \dodoi{10.1016/j.icarus.2016.06.008}

\bibitem[{{Fletcher} {et~al.}(2016{\natexlab{b}}){Fletcher}, {Irwin},
  {Achterberg}, {Orton}, \& {Flasar}}]{16fletcher}
{Fletcher}, L.~N., {Irwin}, P.~G.~J., {Achterberg}, R.~K., {Orton}, G.~S., \&
  {Flasar}, F.~M. 2016{\natexlab{b}}, Icarus, 264, 137,
  \dodoi{10.1016/j.icarus.2015.09.009}

\bibitem[{{Fletcher} {et~al.}(2017{\natexlab{a}}){Fletcher}, {Orton}, {Rogers},
  {Giles}, {Payne}, {Irwin}, \& {Vedovato}}]{17fletcher_seb}
{Fletcher}, L.~N., {Orton}, G.~S., {Rogers}, J.~H., {et~al.}
  2017{\natexlab{a}}, {Icarus}, 286, 94, \dodoi{10.1016/j.icarus.2017.01.001}

\bibitem[{{Fletcher} {et~al.}(2010){Fletcher}, {Orton}, {Mousis},
  {Yanamandra-Fisher}, {Parrish}, {Irwin}, {Fisher}, {Vanzi}, {Fujiyoshi},
  {Fuse}, {Simon-Miller}, {Edkins}, {Hayward}, \& {De Buizer}}]{10fletcher_grs}
{Fletcher}, L.~N., {Orton}, G.~S., {Mousis}, O., {et~al.} 2010, Icarus, 208,
  306, \dodoi{10.1016/j.icarus.2010.01.005}

\bibitem[{{Fletcher} {et~al.}(2011){Fletcher}, {Orton}, {Rogers},
  {Simon-Miller}, {de Pater}, {Wong}, {Mousis}, {Irwin}, {Jacquesson}, \&
  {Yanamandra-Fisher}}]{11fletcher_fade}
{Fletcher}, L.~N., {Orton}, G.~S., {Rogers}, J.~H., {et~al.} 2011, Icarus, 213,
  564, \dodoi{10.1016/j.icarus.2011.03.007}

\bibitem[{{Fletcher} {et~al.}(2017{\natexlab{b}}){Fletcher}, {Orton},
  {Sinclair}, {Donnelly}, {Melin}, {Rogers}, {Greathouse}, {Kasaba},
  {Fujiyoshi}, {Sato}, {Fernandes}, {Irwin}, {Giles}, {Simon}, {Wong}, \&
  {Vedovato}}]{17fletcher_neb}
{Fletcher}, L.~N., {Orton}, G.~S., {Sinclair}, J.~A., {et~al.}
  2017{\natexlab{b}}, Geophys. Res. Lett., 44, 7140,
  \dodoi{10.1002/2017GL073383}

\bibitem[{{Giles} {et~al.}(2015){Giles}, {Fletcher}, \& {Irwin}}]{15giles}
{Giles}, R.~S., {Fletcher}, L.~N., \& {Irwin}, P.~G.~J. 2015, Icarus, 257, 457,
  \dodoi{10.1016/j.icarus.2015.05.030}

\bibitem[{{Giles} {et~al.}(2017{\natexlab{a}}){Giles}, {Fletcher}, \&
  {Irwin}}]{17giles}
---. 2017{\natexlab{a}}, Icarus, 289, 254, \dodoi{10.1016/j.icarus.2016.10.023}

\bibitem[{{Giles} {et~al.}(2017{\natexlab{b}}){Giles}, {Fletcher}, {Irwin},
  {Orton}, \& {Sinclair}}]{17giles_nh3}
{Giles}, R.~S., {Fletcher}, L.~N., {Irwin}, P.~G.~J., {Orton}, G.~S., \&
  {Sinclair}, J.~A. 2017{\natexlab{b}}, Geophys. Res. Lett., 44, 10,
  \dodoi{10.1002/2017GL075221}

\bibitem[{{Grassi} {et~al.}(2017){Grassi}, {Adriani}, {Mura}, {Dinelli},
  {Sindoni}, {Turrini}, {Filacchione}, {Migliorini}, {Moriconi}, {Tosi},
  {Noschese}, {Cicchetti}, {Altieri}, {Fabiano}, {Piccioni}, {Stefani},
  {Atreya}, {Lunine}, {Orton}, {Ingersoll}, {Bolton}, {Levin}, {Connerney},
  {Olivieri}, \& {Amoroso}}]{17grassi}
{Grassi}, D., {Adriani}, A., {Mura}, A., {et~al.} 2017, Geophys. Res. Lett.,
  44, 4615, \dodoi{10.1002/2017GL072841}

\bibitem[{{Hodapp} {et~al.}(2003){Hodapp}, {Jensen}, {Irwin}, {Yamada},
  {Chung}, {Fletcher}, {Robertson}, {Hora}, {Simons}, {Mays}, {Nolan}, {Bec},
  {Merrill}, \& {Fowler}}]{03hodapp}
{Hodapp}, K.~W., {Jensen}, J.~B., {Irwin}, E.~M., {et~al.} 2003, PASP, 115,
  1388, \dodoi{10.1086/379669}

\bibitem[{{Holton} \& {Alexander}(2000)}]{00holton}
{Holton}, J.~R., \& {Alexander}, M.~J. 2000, Washington DC American Geophysical
  Union Geophysical Monograph Series, 123, 21, \dodoi{10.1029/GM123p0021}

\bibitem[{{Hueso} {et~al.}(2017){Hueso}, {S{\'a}nchez-Lavega},
  {I{\~n}urrigarro}, {Rojas}, {P{\'e}rez-Hoyos}, {Mendikoa},
  {G{\'o}mez-Forrellad}, {Go}, {Peach}, {Colas}, \& {Vedovato}}]{17hueso_jup}
{Hueso}, R., {S{\'a}nchez-Lavega}, A., {I{\~n}urrigarro}, P., {et~al.} 2017,
  Geophys. Res. Lett., 44, 4669, \dodoi{10.1002/2017GL073444}

\bibitem[{{Irwin} \& {Dyudina}(2002)}]{02irwin}
{Irwin}, P.~G.~J., \& {Dyudina}, U. 2002, Icarus, 156, 52,
  \dodoi{10.1006/icar.2001.6773}

\bibitem[{{Irwin} {et~al.}(2001){Irwin}, {Weir}, {Taylor}, {Calcutt}, \&
  {Carlson}}]{01irwin}
{Irwin}, P.~G.~J., {Weir}, A.~L., {Taylor}, F.~W., {Calcutt}, S.~B., \&
  {Carlson}, R.~W. 2001, Icarus, 149, 397, \dodoi{10.1006/icar.2000.6542}

\bibitem[{{Kraaikamp}(2016)}]{16kraaikamp}
{Kraaikamp}, E. 2016, {Sky and Telescope}, 132, 68

\bibitem[{{Lagage} {et~al.}(2004){Lagage}, {Pel}, {Authier}, {Belorgey},
  {Claret}, {Doucet}, {Dubreuil}, {Durand}, {Elswijk}, {Girardot}, {K{\"a}ufl},
  {Kroes}, {Lortholary}, {Lussignol}, {Marchesi}, {Pantin}, {Peletier},
  {Pirard}, {Pragt}, {Rio}, {Schoenmaker}, {Siebenmorgen}, {Silber}, {Smette},
  {Sterzik}, \& {Veyssiere}}]{04lagage}
{Lagage}, P.~O., {Pel}, J.~W., {Authier}, M., {et~al.} 2004, The Messenger,
  117, 12

\bibitem[{{Li} {et~al.}(2006){Li}, {Ingersoll}, {Vasavada}, {Simon-Miller},
  {Achterberg}, {Ewald}, {Dyudina}, {Porco}, {West}, \& {Flasar}}]{06li}
{Li}, L., {Ingersoll}, A.~P., {Vasavada}, A.~R., {et~al.} 2006, Icarus, 185,
  416, \dodoi{10.1016/j.icarus.2006.08.005}

\bibitem[{{Nixon} {et~al.}(2001){Nixon}, {Irwin}, {Calcutt}, {Taylor}, \&
  {Carlson}}]{01nixon}
{Nixon}, C.~A., {Irwin}, P.~G.~J., {Calcutt}, S.~B., {Taylor}, F.~W., \&
  {Carlson}, R.~W. 2001, Icarus, 150, 48, \dodoi{10.1006/icar.2000.6561}

\bibitem[{{Rogers}(2017)}]{17rogers}
{Rogers}, J.~H. 2017, Journal of the British Astronomical Association.
\newblock \doarXiv{1707.03343}

\bibitem[{{Rogers} {et~al.}(2016){Rogers}, {Fletcher}, {Adamoli}, {Jacquesson},
  {Vedovato}, \& {Orton}}]{16rogers_wave}
{Rogers}, J.~H., {Fletcher}, L.~N., {Adamoli}, G., {et~al.} 2016, Icarus, 277,
  354.
\newblock \doarXiv{1605.07883}

\bibitem[{{S{\'a}nchez-Lavega} {et~al.}(2011){S{\'a}nchez-Lavega}, {del
  R{\'{\i}}o-Gaztelurrutia}, {Hueso}, {G{\'o}mez-Forrellad}, {Sanz-Requena},
  {Legarreta}, {Garc{\'{\i}}a-Melendo}, {Colas}, {Lecacheux}, {Fletcher},
  {Barrado y Navascu{\'e}s}, {Parker}, {International Outer Planet Watch Team},
  {Akutsu}, {Barry}, {Beltran}, {Buda}, {Combs}, {Carvalho}, {Casquinha},
  {Delcroix}, {Ghomizadeh}, {Go}, {Hotershall}, {Ikemura}, {Jolly}, {Kazemoto},
  {Kumamori}, {Lecompte}, {Maxson}, {Melillo}, {Milika}, {Morales}, {Peach},
  {Phillips}, {Poupeau}, {Sussenbach}, {Walker}, {Walker}, {Tranter}, {Wesley},
  {Wilson}, \& {Yunoki}}]{11sanchez}
{S{\'a}nchez-Lavega}, A., {del R{\'{\i}}o-Gaztelurrutia}, T., {Hueso}, R.,
  {et~al.} 2011, Nature, 475, 71, \dodoi{10.1038/nature10203}

\bibitem[{{S{\'a}nchez-Lavega} {et~al.}(2017){S{\'a}nchez-Lavega}, {Rogers},
  {Orton}, {Garc{\'{\i}}a-Melendo}, {Legarreta}, {Colas}, {Dauvergne}, {Hueso},
  {Rojas}, {P{\'e}rez-Hoyos}, {Mendikoa}, {I{\~n}urrigarro}, {Gomez-Forrellad},
  {Momary}, {Hansen}, {Eichstaedt}, {Miles}, \& {Wesley}}]{17sanchez_NTB}
{S{\'a}nchez-Lavega}, A., {Rogers}, J.~H., {Orton}, G.~S., {et~al.} 2017,
  Geophys. Res. Lett., 44, 4679, \dodoi{10.1002/2017GL073421}

\bibitem[{{Simon} {et~al.}(2015){Simon}, {Wong}, \& {Orton}}]{15simon}
{Simon}, A.~A., {Wong}, M.~H., \& {Orton}, G.~S. 2015, ApJ Letters, 812, 55,
  \dodoi{10.1088/0004-637X/812/1/55}

\bibitem[{{Simon} {et~al.}(2018){Simon}, {Hueso}, {Inurrigarro},
  {Sanchez-Lavega}, {Morales-Juber{\'{\i}}as}, {Cosentino}, {Fletcher}, {Wong},
  {de Pater}, {Orton}, {Colas}, {Delcroix}, {Peach}, \&
  {Gomez-Forrellad}}]{18simon}
{Simon}, A.~A., {Hueso}, R., {Inurrigarro}, P., {et~al.} 2018, AJ, accepted

\bibitem[{{Smith} {et~al.}(1979){Smith}, {Soderblom}, {Johnson}, {Ingersoll},
  {Collins}, {Shoemaker}, {Hunt}, {Masursky}, {Carr}, {Davies}, {Cook},
  {Boyce}, {Owen}, {Danielson}, {Sagan}, {Beebe}, {Veverka}, {McCauley},
  {Strom}, {Morrison}, {Briggs}, \& {Suomi}}]{79smith}
{Smith}, B.~A., {Soderblom}, L.~A., {Johnson}, T.~V., {et~al.} 1979, Science,
  204, 951, \dodoi{10.1126/science.204.4396.951}

\bibitem[{{Sromovsky} \& {Fry}(2010)}]{10sromovsky}
{Sromovsky}, L.~A., \& {Fry}, P.~M. 2010, Icarus, 210, 230,
  \dodoi{10.1016/j.icarus.2010.06.039}

\bibitem[{{Terrile} \& {Westphal}(1977)}]{77terrile}
{Terrile}, R.~J., \& {Westphal}, J.~A. 1977, Icarus, 30, 274,
  \dodoi{10.1016/0019-1035(77)90159-2}

\end{thebibliography}
\end{document}